\documentclass[floatfix,aps,amsmath,prd,twocolumn,eqsecnum,showpacs,nofootinbib,reprint]{revtex4-1}
\usepackage{graphicx,graphics,amsfonts,amsmath,amssymb,bm,hyperref,psfrag,natbib,dcolumn}

\allowdisplaybreaks

\newcommand{\be}{\begin{equation}}
\newcommand{\ee}{\end{equation}}
\newcommand{\bs}{\begin{subequations}}
\newcommand{\es}{\end{subequations}}

\DeclareMathOperator{\sign}{sign}

\begin{document}
\title{Conservative corrections to the innermost stable circular orbit (ISCO) of a Kerr black hole: a new gauge-invariant post-Newtonian ISCO condition, and the ISCO shift due to test-particle spin and the gravitational self-force}
\author{Marc Favata}
\thanks{NASA Postdoctoral Fellow}
\email{favata@tapir.caltech.edu}
\affiliation{Jet Propulsion Laboratory, 4800 Oak Grove Drive, Pasadena, California 91109, USA}
\thanks{Copyright 2010 California Institute of Technology. Government sponsorship acknowledged.}
\affiliation{Theoretical Astrophysics, 350-17, California Institute of Technology, Pasadena, California 91125, USA}
\date{12 October 2010}
\begin{abstract}
The innermost stable circular orbit (ISCO) delimits the transition from circular orbits to those that plunge into a black hole. In the test-mass limit, well-defined ISCO conditions exist for the Kerr and Schwarzschild spacetimes. In the finite-mass case, there are a large variety of ways to define an ISCO in a post-Newtonian (PN) context. Here I generalize the gauge-invariant ISCO condition of Blanchet and Iyer [Classical Quantum Gravity {\bf 20}, 755 (2003)] to the case of spinning (nonprecessing) binaries. The Blanchet-Iyer ISCO condition has two desirable and unexpected properties: (1) it exactly reproduces the Schwarzschild ISCO in the test-mass limit, and (2) it accurately approximates the recently calculated shift in the Schwarzschild ISCO frequency due to the conservative-piece of the gravitational self-force [Barack and Sago, Phys.~Rev.~Lett.~{\bf 102}, 191101 (2009)]. The generalization of this ISCO condition to spinning binaries has the property that it also exactly reproduces the Kerr ISCO in the test-mass limit (up to the order at which PN spin corrections are currently known). The shift in the ISCO due to the spin of the test-particle is also calculated. Remarkably, the gauge-invariant PN ISCO condition exactly reproduces the ISCO shift predicted by the Papapetrou equations for a fully-relativistic spinning particle. It is surprising that an analysis of the stability of the standard PN equations of motion is able (without any form of ``resummation'') to accurately describe strong-field effects of the Kerr spacetime. The ISCO frequency shift due to the conservative self-force in Kerr is also calculated from this new ISCO condition, as well as from the effective-one-body Hamiltonian of Barausse and Buonanno [Phys.~Rev.~D {\bf 81}, 084024 (2010)]. These results serve as a useful point of comparison for future gravitational self-force calculations in the Kerr spacetime.
\end{abstract}
\pacs{04.25.Nx, 04.25.-g, 04.25.D-, 04.30.-w}
\maketitle
\section{\label{sec:intro}Introduction, Motivation, and Summary}
The innermost stable circular orbit (ISCO) is a point of dynamical instability in black hole (BH) spacetimes that separates stable, circular, and bound geodesic orbits from those that ``plunge'' into the BH event horizon. The location of the ISCO can be quantified in a gauge-invariant manner by specifying its orbital angular frequency as measured by a distant observer. For a test-particle in the Schwarzschild spacetime, this frequency occurs at $m_2 \Omega = 6^{-3/2}$, where $m_2$ is the mass of the BH.\footnote{Throughout this article $m_1<m_2$ denote the binary masses, $q=m_1/m_2\leq 1$ is the mass ratio, $M=m_1+m_2$ is the total mass, and $\eta=m_1 m_2/M^2=q/(1+q)^2\leq 1/4$ is the reduced mass ratio (denoted $\nu$ by some authors).} The location of the ISCO is important in the context of quasicircular, inspiralling compact binaries (an important source for ground and space-based gravitational-wave detectors) because it represents the point where the character of the orbit (and hence the gravitational waves) abruptly changes. Because of this, the ISCO frequency is often taken as the termination point of inspiral templates. The ISCO is also important because its location encodes (potentially observable) information about the strong-gravity region of the BH spacetime.

What happens if we no longer have a geodesic orbit? When dissipation (i.e., radiation-reaction) is included, the location of the ISCO is no longer precisely quantifiable---it becomes ``blurred'' into a transition region (in orbital radius or frequency) separating the adiabatic inspiral from the plunge \cite{ot00,EOB-BD2}. However, if we consider only conservative corrections to geodesic motion, a precise ISCO can (in some cases) continue to exist. In particular here we will consider two types of conservative corrections to geodesic motion: (i) the \emph{gravitational self-force} (GSF; a force arising from the point-particle's finite mass which causes it to deviate from geodesic motion) and (ii) the force due to the spin of the test-body.\footnote{The quadrupole and higher-order multipole moments of an extended body could also cause a shift in the ISCO location.}

Calculations of the GSF are motivated by the need to model extreme-mass-ratio inspirals (EMRIs), an important source for the planned Laser Interferometer Space Antenna (LISA) \cite{lisaweb} consisting of a compact object ($m_1\sim 1 \mbox{--} 100 M_{\odot}$) inspiralling into a massive BH ($m_2 \sim 10^4 \mbox{--} 10^7 M_{\odot}$) with mass ratios $q \lesssim10^{-4}$. Computing the GSF is challenging (see \cite{poissonselforcereview,lousto-selfforcereview,detweiler-selfforcelec,barack-selfforcereview} for reviews and references), but several groups have had recent success \cite{barack-sago-circselfforcePRD2007,detweiler-circselfforcePRD2008,berndtson-selfforcePhDthesis,sago-barack-detweiler,keidl-etal-GSFkerrI,shah-etal-conservGSF}.
In particular, one of the concrete results to emerge from the self-force program has been the calculation by Barack and Sago (BS) of the shift in the ISCO frequency due to the conservative GSF in the Schwarzschild spacetime \cite{barack-sago_isco,barack-sago-eccentricselfforce}. This result is especially interesting because it supplies a gauge-invariant, exact strong-field result that is only computable using the full self-force formalism. (This is in contrast to standard BH perturbation theory calculations, which only provide access to the time-averaged dissipative pieces of the self-force.) The resulting conservative GSF ISCO frequency shift can be expressed in the form
\be
M \Omega = 6^{-3/2} [1+ \eta c^{\rm GSF}(0) + O(\eta^2)],
\ee
where BS calculated the value $c^{\rm GSF}(0)=1.2512 (\pm 0.0004)$. This value can be used to compare different GSF codes, and to set constraints on the effective-one-body (EOB) \cite{EOB-BD1,EOB-BD2,EOB-damour-lecnotes,damour-nagar-EOBlecnotes2009} formalism (see \cite{damour-GSF,barausse-buonanno-spinEOB,favata-iscostudy}).

In Ref.~\cite{favata-iscostudy}  I compared the above GSF ISCO shift with $\sim 15$ distinct post-Newtonian (PN) or EOB methods for computing the ISCO. Among those methods, two approaches---based on the EOB formalism and the standard PN equations of motion---have especially desirable features. In particular, the best agreement ($\sim 10\%$ error) with the BS result was found using a version of the EOB formalism in which a pseudo-4PN term is added to the effective metric and calibrated with the Caltech/Cornell numerical relativity simulations \cite{buonanno-caltechEOB09}. This method also adequately predicted (with $\sim 16\%$ error) the ISCO frequency for equal-mass binaries as computed from sequences of quasicircular initial data \cite{caudill-etal-initialdata-PRD2006}. However, in the absence of calibration, the method which most accurately reproduced the BS result was the gauge-invariant ISCO condition of Blanchet and Iyer \cite{blanchetiyer3PN}.\footnote{In addition to the above-mentioned reasons, these two methods for computing the ISCO are also preferred over the other approaches examined in \cite{favata-iscostudy} because: (i) the error in the ISCO computed via these methods decreases monotonically as the PN order is increased, and (ii) they each are derived from equations of motion that allow for a complete description of the two-body dynamics.} This condition is derived from a stability analysis of the 3PN (nonspinning) equations of motion; it takes the form
\begin{multline}
\label{eq:C0invar}
\hat{C}_0 \equiv 1-6x + 14\eta x^2
\\
+ \left[ \left(\frac{397}{2} -\frac{123}{16}\pi^2\right)\eta -14\eta^2\right] x^3 + O(x^4),
\end{multline}
where $x\equiv (M\Omega)^{2/3}$, and $\hat{C}_0 \geq 0$ is required for stable, circular orbits to exist. The ISCO is found by solving $\hat{C}_0=0$ for $x$ (or $\Omega$). The resulting value for the conservative GSF ISCO shift was found to be \cite{favata-iscostudy}
\be
\label{eq:c3PNvalues}
c_{C_0}^{\rm GSF}(0) \equiv \frac{565}{288} - \frac{41\pi^2}{768} = 1.434\,912\,612\ldots,
\ee
which differs from the exact BS result by $14.7\%$.

The above PN ISCO condition is especially interesting because it exactly reproduces the Schwarzschild ISCO ($x=1/6$ or $m_2\Omega=6^{-3/2}$) in the test-particle limit. It is surprising that a condition derived from the PN equations of motion can reproduce a strong-field result like the ISCO.\footnote{Indeed, one can see from Eq.~\eqref{eq:C0invar} that the Schwarzschild ISCO frequency arises only from the 1PN equations of motion; the 2PN and 3PN terms affect only the $O(\eta)$ corrections. Note also that in deriving this result, it was crucial to express $\hat{C}_0$ in terms of the gauge-invariant observable $x$ rather than a gauge-dependent radial coordinate \cite{blanchetiyer3PN}.} For example, a standard way to compute the ISCO in a PN context is by finding the minimum of the circular-orbit energy.\footnote{The critical point defined in this way is sometimes called an ICO (innermost circular orbit). See Sec.~\mbox{II~B} of \cite{favata-iscostudy} (as well as Sec.~\mbox{IV~A~2} of \cite{bcv1,*bcv1-erratum}) for a discussion of the difference and relationship between the ISCO and ICO. In the rest of this article, I will refer to both terms as an ISCO.} In the test-mass limit, the PN expansion of the circular-orbit energy,
\begin{multline}
\frac{E_{\rm circ}(\Omega)}{\eta M} = \frac{(1-2x)}{(1-3x)^{1/2}} -1 = -\frac{1}{2} x \left[ 1 -\frac{3}{4} x
-\frac{27}{8} x^2
\right.
\\ \left.
-\frac{675}{64} x^3 -\frac{3969}{128} x^4 -\frac{45\,927}{512} x^5 + O(x^6) \right],
\end{multline}
converges slowly: to get within $8\%$ of the exact result ($x=1/6)$ one needs to truncate the above expression at 4PN order or higher.

Part of the motivation for developing ``resummation'' methods was to cure this problem while also providing a means to compute the ISCO for finite mass-ratio binaries. For example, Kidder, Will, and Wiseman \cite{kidderwillwiseman-transition-CQG1992,kidderwillwiseman-transition-PRD1993} modified the PN equations of motion by replacing the $O(\eta^0)$ terms with the corresponding terms derived from the Schwarzschild geodesic equations (in the appropriate coordinate system). This enforced the Schwarzschild ISCO in the test-particle limit, but caused deviations from this value for finite-$\eta$. Similarly, Ref.~\cite{damour-iyer-sathyaprakash-PRD1998} introduced Pad\'{e} approximants to improve the convergence of PN-based templates (in part by again enforcing agreement with the test-particle limit). The EOB formalism provides the most successful version of this idea by modeling the two-body dynamics in terms of a Hamiltonian that is based on a particle with reduced mass $\mu=\eta M$ moving in the ``$\eta$-deformed'' Schwarzschild background of a central mass $M$. It is in light of these resummation approaches that the ability of the Blanchet-Iyer ISCO condition to predict the Schwarzschild ISCO is surprising (and perhaps not widely appreciated).

\subsection{\label{sec:summary}Summary of results}
It is possible that the ability of the Blanchet-Iyer ISCO condition to predict the Schwarzschild ISCO is coincidental. One of the primary objectives of this study is to test this by extending the Blanchet-Iyer ISCO condition [Eq.~\eqref{eq:C0invar}] to the case of spinning (nonprecessing) binaries. This calculation is performed in Sec.~\ref{sec:C0spinning}. The result is given by [see also Eq.~\eqref{eq:C0hat-spin} below]
\begin{multline}
\label{eq:C0hat-spin-intro}
\hat{C}_0 \equiv  1- 6x + x^{3/2} \left( 14 \frac{S^{\rm c}_{\ell}}{M^2} + 6 \frac{\delta m}{M} \frac{\Sigma^{\rm c}_{\ell}}{M^2} \right)
\\
+ x^2 \left[ 14\eta - 3 \left( \frac{S^{\rm c}_{0,\ell}}{M^2} \right)^2 \right]
\\
+ x^{5/2} \left[ - \frac{S_{\ell}^{\rm c}}{M^2} (22+32\eta) - \frac{\delta m}{M} \frac{\Sigma_{\ell}^{\rm c}}{M^2} (18+15\eta) \right]
\\
+ x^3 \left[ \left( \frac{397}{2} - \frac{123}{16}\pi^2  \right) \eta - 14 \eta^2 \right],
\end{multline}
where $S^{\rm c}_{\ell} \equiv {\bm \ell} \cdot {\mathbf S}^{\rm c}$, $\Sigma^{\rm c}_{\ell} \equiv {\bm \ell} \cdot {\mathbf \Sigma}^{\rm c}$, $S^{\rm c}_{0,\ell} \equiv {\bm \ell} \cdot {\mathbf S}_0^{\rm c}$, ${\bm \ell}$ is the unit vector along the direction of the Newtonian orbital angular momentum, ${\mathbf S}^{\rm c} \equiv {\mathbf S}^{\rm c}_1 + {\mathbf S}^{\rm c}_2$, ${\mathbf \Sigma}^{\rm c} \equiv M ({\mathbf S}_2^{\rm c}/m_2 - {\mathbf S}_1^{\rm c}/m_1)$, ${\mathbf S}^{\rm c}_0 = (1+ m_2/m_1) {\mathbf S}_1^{\rm c} + (1+ m_1/m_2 ) {\mathbf S}_2^{\rm c}$, $\delta m = m_1 - m_2$, and ${\mathbf S}_A^{\rm c} = \chi_A^{\rm c} m_A^2 {\hat{\mathbf s}}_A^{\rm c}$ are the individual spin angular momenta for body $A=1,2$ with dimensionless spin parameters $\chi_A^{\rm c}$ and unit direction vectors ${\hat{\mathbf s}}_A^{\rm c}$. This condition is derived from the 3PN equations of motion, including all explicitly known spin terms up to 2.5PN order.

\begin{figure}[t]
\includegraphics[angle=0, width=0.48\textwidth]{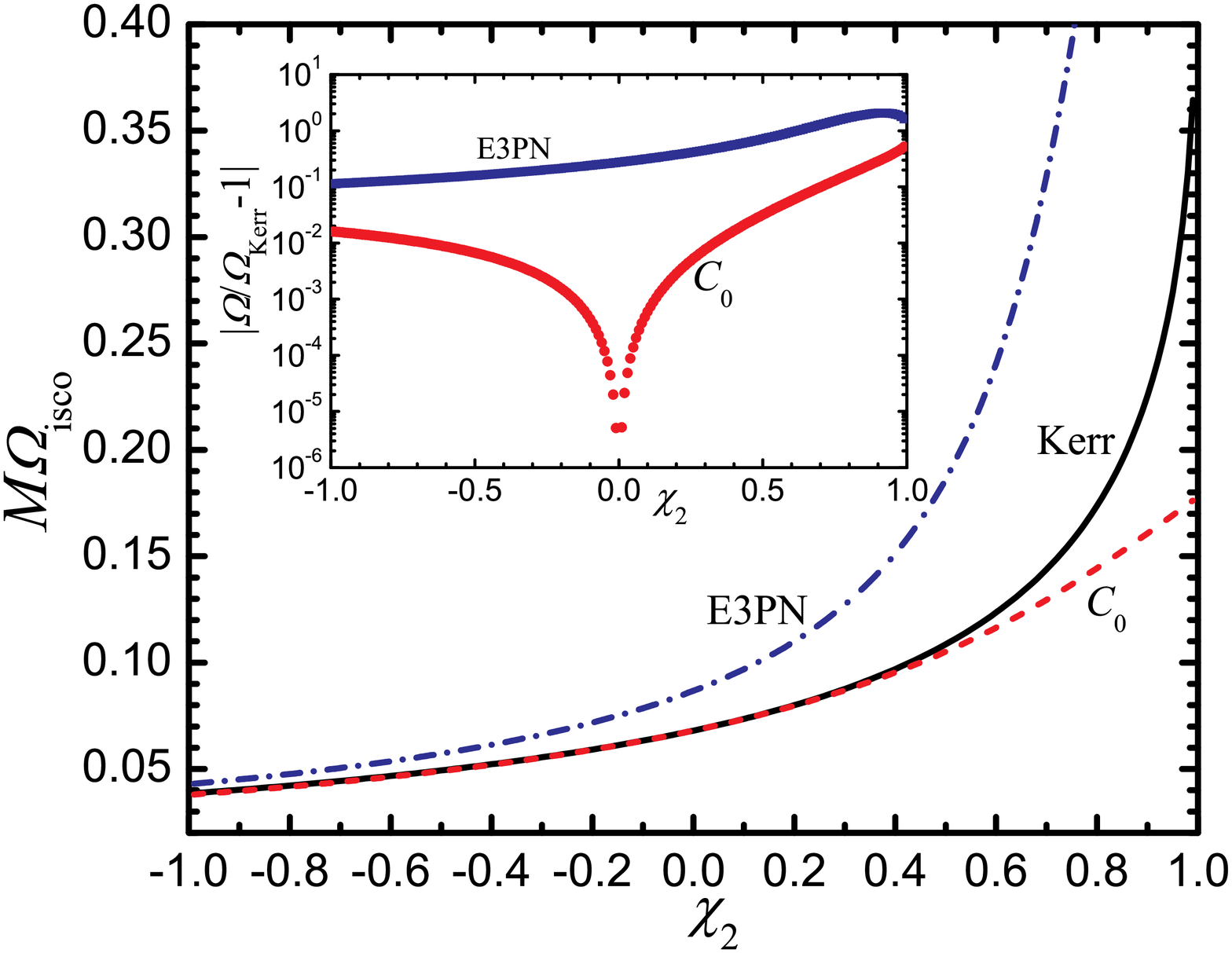}
\caption{\label{fig:isco-compare}(color online). Comparison of three different methods for computing the ISCO of a nonspinning test-particle in the Kerr spacetime. The solid (black) curve (labeled Kerr) refers to the exact result for the Kerr ISCO [Ref.~\cite{bpt} or Eq.~\eqref{eq:C0kerr} here]. The dashed (red) curve (labeled $C_0$) is the $\eta\rightarrow 0$ limit of the gauge-invariant ISCO condition derived here [Eq.~\eqref{eq:C0hat-spin-intro} or \eqref{eq:C0hat-spin}]. The dash-dotted (blue) curve (labeled E3PN) is the ISCO computed by minimizing the 3PN circular-orbit energy [Eq.~\eqref{eq:E3PN}]. The inset shows the fractional errors of the E3PN or $C_0$ curves with respect to the Kerr curve.}
\end{figure}

In the test-particle limit ($\eta \rightarrow 0$), the ISCO determined from Eq.~\eqref{eq:C0hat-spin-intro} can be compared with the ISCO of the Kerr spacetime \cite{bpt}. This comparison can be performed by deriving a condition analogous to $\hat{C}_0$ from the Kerr metric, expanding the result in powers of the BH spin ($\chi_2^{\rm K}$), and comparing to Eq.~\eqref{eq:C0hat-spin-intro} (see Sec.~\ref{sec:kerriscocompare} for details). The resulting comparison shows that the two conditions agree up to the order to which the PN spin corrections are known. This comparison is also shown graphically in Fig.~\ref{fig:isco-compare}. Note the large improvement in comparison with the 3PN energy function [which includes spin corrections; see Eq.~\eqref{eq:E3PN}]. Presumably, if higher-order spin corrections in the PN equations of motion were included, the error in comparison with the Kerr ISCO for large values of $|\chi_2|$ would improve. This excellent agreement suggests that the standard PN equations of motion are able to exactly recover some strong-field results.

A second objective of this article is to calculate the shift in the ISCO frequency due to conservative effects (Sec.~\ref{sec:ISCOshift}). In particular, two types of conservative effects are considered: the first due to the GSF, and the second due to the spin of the test-particle. As discussed above, the conservative GSF ISCO shift was computed in \cite{barack-sago_isco,barack-sago-eccentricselfforce} for a Schwarzschild background, and compared with various PN calculations in \cite{favata-iscostudy}. Here we focus on the ISCO shift in the Kerr background, for which GSF calculations are not currently available. Instead, we make predictions for what that ISCO shift might be according to two analytic approaches: the ISCO condition in Eq.~\eqref{eq:C0hat-spin-intro} above and the recently developed spinning-EOB formalism of Barausse and Buonanno \cite{barausse-buonanno-spinEOB} (Sec.~\ref{sec:spinEOB}). (In the latter case, the ISCO shift was calibrated to match the exact Schwarzschild result \cite{barack-sago_isco,barack-sago-eccentricselfforce}.)

To quantify these conservative ISCO shifts, we expand the ISCO frequency as [Eq.~\eqref{eq:ISCO-expand-schematic3} below]
\begin{multline}
\label{eq:ISCO-expand-schematic3-intro}
M\Omega = m_2 \Omega^{\rm K}(\chi_2) [ 1+ \eta c^{\rm GSF}(\chi_2) + \eta \chi_1 c^{\rm COspin}(\chi_2)
\\
+ O(\eta^2) + O(\chi_1 \eta^2) +O(\chi_1^2 \eta^2)],
\end{multline}
where $\Omega^{\rm K}(\chi_2)$ is the Kerr ISCO frequency \cite{bpt}. The shift in the ISCO due to the conservative GSF is parametrized by the function $c^{\rm GSF}(\chi_2)$. In Sec.~\ref{sec:cGSF} this function is calculated via the EOB and $\hat{C}_0$ approaches; the results are presented graphically in Fig.~\ref{fig:cEOB-PN} and tabulated in Table \ref{tab:isco-quantities}. It will be interesting to compare these numbers with future GSF calculations in Kerr.

In Sec.~\ref{sec:cCOspin} the function $c^{\rm COspin}(\chi_2)$ is also calculated via the EOB and $\hat{C}_0$ approaches.\footnote{The ISCO for a spinning test-particle in Kerr was previously considered in \cite{suzuki-maeda-COspinISCO}, but those authors focused on unphysically large values of the test-particle spin and did not explicitly compute the shift parameter $c^{\rm COspin}$ (see also \cite{tanaka-mino-sasaki-shibata-PRD1996-spinningparticlePN,suzuki-maeda-PRD1997-chaos-schw}).} However, in this case the EOB calculation via the Hamiltonian in \cite{barausse-buonanno-spinEOB} yields the exact (fully relativistic) result. This is because this Hamiltonian reproduces the Papapetrou-Mathisson-Dixon equations of motion \cite{papapetrou-spinningI-1951,papapetrou-spinningII-1951,mathisson-1931-original,*mathisson-1931-reprint,mathisson-1937-original,*mathisson-1937-reprint,dixon-extendedbodiesI-PRSocA1970,dixon-extendedbodiesII-PRSocA1970,dixon-extendedbodiesIII-PRSocA1974,barausse-racine-buonanno-PRD2009} in the small-$\eta$ limit.
An analysis of the ISCO shift directly using the Papapetrou equations is also presented in Appendix \ref{app:papa}; the results are identical to those obtained from the EOB Hamiltonian (providing further confirmation of the work in \cite{barausse-racine-buonanno-PRD2009,barausse-buonanno-spinEOB}).
These results are shown in Fig.~\ref{fig:cEOB-PN} and Table \ref{tab:isco-quantities}. In the Schwarzschild case a fully analytic analysis of the Papapetrou equations is straightforward and presented in Appendix \ref{subsec:Papaschw}. It shows that the Boyer-Lindquist radial coordinate of the Schwarzschild ISCO is shifted by $O(\chi_1 m_1)$:
\be
r_{\rm isco} = 6 m_2 - 2 \sqrt{\frac{2}{3}} \chi_1 m_1 + O(\chi_1^2 m_1^2),
\ee
and frequency shift of the ISCO due to the point-particle spin is given by
\be
\label{eq:C0spin-schw-into}
c^{\rm COspin}(0) = \frac{\sqrt{6}}{8} = 0.306\,186\,\ldots.
\ee
Interestingly, this ISCO frequency shift is exactly reproduced by the $\hat{C}_0$ ISCO condition, again showing that the standard PN equations of motion are able to exactly reproduce a strong-field result. (If $\chi_2 \neq 0$, the exact result is only approximately reproduced by the $\hat{C}_0$ condition because the PN spin terms are explicitly computed only to 2.5PN order; see Fig.~\ref{fig:cEOB-PN}.)

Section \ref{sec:conclusion} discusses some conclusions of this study. Appendix \ref{app:spincheck} compares the test-mass limits of several PN quantities (the orbital energy, angular momentum, and Keplerian relation) with the analogous quantities computed from the Kerr metric.

\section{\label{sec:C0spinning}Gauge-invariant ISCO condition for spinning binaries}
Following the stability analysis of the PN equations of motion in \cite{kidderwillwiseman-transition-CQG1992,blanchetiyer3PN}, we can generalize the gauge-invariant ISCO condition derived by Blanchet and Iyer \cite{blanchetiyer3PN} to the case of spinning, nonprecessing binaries.

We begin by writing the conservative PN equations of motion for two spinning point-masses as
\begin{multline}
\label{eq:PNspineqns}
\frac{d{\mathbf v}}{dt} = \mathop{\mathbf{B}}_\text{NS}{}^{\!\!\mathrm{N}}  + \mathop{\mathbf{B}}_\text{NS}{}^{\!\!\mathrm{1PN}}  + \mathop{\mathbf{B}}_\text{NS}{}^{\!\!\mathrm{2PN}}  + \mathop{\mathbf{B}}_\text{NS}{}^{\!\!\mathrm{3PN}}  \\
+ \mathop{\mathbf{B}}_\text{SO}{}^{\!\!\mathrm{1.5PN}}  + \mathop{\mathbf{B}}_\text{SO}{}^{\!\!\mathrm{2.5PN}}  + \mathop{\mathbf{B}}_\text{SS+QM}{}^{\!\!\!\!\!\!\!\!\mathrm{2PN}} .
\end{multline}
On the first line we list the nonspin terms to 3PN order (see \cite{blanchetLRR} for references); note that the radiation-reaction terms at 2.5PN and 3.5PN order are not present since we are only concerned with conservative corrections to the ISCO. The spin-orbit (SO) term at 1.5PN order and the spin-spin (SS) term at 2PN order were first derived in \cite{barker-oconnell-PRD1970}. The SO term at 2.5PN order was first derived in \cite{tagoshi-ohashi-owen-2.5SOterm}. Here I use the forms given in Eqs.~(5.7) of \cite{faye-buonanno-luc-higherorderspinI}. The 2PN order quadrupole-monopole (QM) term was derived in \cite{poisson-quadrupolemonopoleterm-PRD1998}; Ref.~\cite{racine-buonanno-kidder-spinningrecoil} shows how to concisely combine this term (when specialized to black holes) with the 2PN order spin-spin term [see their Eq.~(3.8)].
\subsection{\label{sec:spinvars}Equations of motion and the relationship between spin variables}
The spin-orbit contributions to the equations of motion given in \cite{faye-buonanno-luc-higherorderspinI} are expressed in terms of spin vectors ${\mathbf S}^{\rm nc}_A$ ($A=1,2$) whose magnitudes $\chi_A^{\rm nc} m_A^2$ \emph{do not} remain constant with time. (Note that Refs.~\cite{faye-buonanno-luc-higherorderspinI,faye-buonanno-luc-higherorderspinII,*faye-buonanno-luc-higherorderspinIIerratum,*faye-buonanno-luc-higherorderspinIIerratum2} do not use the superscripts ``${\rm nc}$''.) An alternative set of spin variables ${\mathbf S}_A^{\rm c}$ are defined in Eq.~(7.4) of \cite{faye-buonanno-luc-higherorderspinII,*faye-buonanno-luc-higherorderspinIIerratum,*faye-buonanno-luc-higherorderspinIIerratum2} [also Eq.~(2.21) of \cite{racine-buonanno-kidder-spinningrecoil}] and have the property that their magnitudes \emph{are} constant.\footnote{Throughout this section all of our spin variables are contravariant vectors. In \cite{racine-buonanno-kidder-spinningrecoil} these are denoted with an overbar. Note that the spin variables used in Kidder \cite{kidder-spineffects} are the constant-magnitude, contravariant spin vectors denoted ${\mathbf S}_A^{\rm c}$ here. Note also that we use the notation ${\mathbf \Sigma}$ for the quantities denoted ${\mathbf \Delta}$ in \cite{kidder-spineffects,racine-buonanno-kidder-spinningrecoil}.} This choice of spin variables causes the spin-precession equations to take a convenient form and is generally preferred in computations. Here we denote the nonconstant-magnitude spin vectors of each body by ${\mathbf S}^{\rm nc}_A$, and the constant-magnitude spin vectors by ${\mathbf S}^{\rm c}_A$.  We also define the spin combinations
\bs
\begin{align}
{\mathbf S}^{\rm c} &\equiv {\mathbf S}^{\rm c}_1 + {\mathbf S}^{\rm c}_2, \\
{\mathbf \Sigma}^{\rm c} &\equiv M \left( \frac{{\mathbf S}^{\rm c}_2}{m_2} - \frac{{\mathbf S}^{\rm c}_1}{m_1} \right),
\end{align}
\es
and analogous relations for ${\mathbf S}^{\rm nc}$ and ${\mathbf \Sigma}^{\rm nc}$.

The relationship between $({\mathbf S}^{\rm nc}, {\mathbf \Sigma}^{\rm nc})$ and $({\mathbf S}^{\rm c}, {\mathbf \Sigma}^{\rm c})$ is given by Eqs.~(2.22) of \cite{racine-buonanno-kidder-spinningrecoil},
\bs
\label{eq:spintransform-racine}
\begin{multline}
{\mathbf S}^{\rm c} = {\mathbf S}^{\rm nc} + \frac{1}{c^2} \left\{ \eta \frac{M}{r} \left[ 2 {\mathbf S}^{\rm nc} + \frac{\delta m}{M} {\mathbf \Sigma}^{\rm nc} \right]
\right. \\ \left.
- \frac{\eta}{2} \left[ {\mathbf v} \cdot {\mathbf S}^{\rm nc} + \frac{\delta m}{M} {\mathbf v} \cdot {\mathbf \Sigma}^{\rm nc} \right] {\mathbf v} \right\} + O(c^{-4}),
\end{multline}
\begin{multline}
{\mathbf \Sigma}^{\rm c} = {\mathbf \Sigma}^{\rm nc} + \frac{1}{c^2} \left\{ \frac{M}{r} \left[ \frac{\delta m}{M} {\mathbf S}^{\rm nc} + (1-2\eta) {\mathbf \Sigma}^{\rm nc} \right]
\right. \\ \left.
- \frac{1}{2} \left[ \frac{\delta m}{M} {\mathbf v} \cdot {\mathbf S}^{\rm nc} + (1-3\eta) {\mathbf v} \cdot {\mathbf \Sigma}^{\rm nc} \right] {\mathbf v} \right\} + O(c^{-4}),
\end{multline}
\es
where $r$ is the orbital separation in harmonic coordinates.
The inverse relationship is given by
\bs
\label{eq:spintransform}
\begin{multline}
{\mathbf S}^{\rm nc} = {\mathbf S}^{\rm c} + \frac{1}{c^2} \left\{ -\eta \frac{M}{r} \left[ 2 {\mathbf S}^{\rm c} + \frac{\delta m}{M} {\mathbf \Sigma}^{\rm c} \right]
\right. \\ \left.
+ \frac{\eta}{2} \left[ {\mathbf v} \cdot {\mathbf S}^{\rm c} + \frac{\delta m}{M} {\mathbf v} \cdot {\mathbf \Sigma}^{\rm c} \right] {\mathbf v} \right\} + O(c^{-4}),
\end{multline}
\begin{multline}
{\mathbf \Sigma}^{\rm nc} = {\mathbf \Sigma}^{\rm c} + \frac{1}{c^2} \left\{ -\frac{M}{r} \left[ \frac{\delta m}{M} {\mathbf S}^{\rm c} + (1-2\eta) {\mathbf \Sigma}^{\rm c} \right]
\right. \\ \left.
+ \frac{1}{2} \left[ \frac{\delta m}{M} {\mathbf v} \cdot {\mathbf S}^{\rm c} + (1-3\eta) {\mathbf v} \cdot {\mathbf \Sigma}^{\rm c} \right] {\mathbf v} \right\} + O(c^{-4}),
\end{multline}
\es
where the powers of $c$ were added to show that the corrections to the spins are a relative 1PN order effect.
We also note the relationship between the individual spin vectors \cite{racine-buonanno-kidder-spinningrecoil,faye-buonanno-luc-higherorderspinI,faye-buonanno-luc-higherorderspinII,*faye-buonanno-luc-higherorderspinIIerratum,*faye-buonanno-luc-higherorderspinIIerratum2},
\bs
\label{eq:relatespins}
\begin{align}
{\mathbf S}^{\rm c}_A &= \left(1+ \frac{m_B}{c^2 r}\right) {\mathbf S}^{\rm nc}_A - \frac{1}{2 c^2} \left( \frac{m_B}{M}\right)^2 ({\mathbf v} \cdot {\mathbf S}^{\rm nc}_A) {\mathbf v} + O(c^{-4}),\\
{\mathbf S}^{\rm nc}_A &= \left(1 - \frac{m_B}{c^2 r}\right) {\mathbf S}^{\rm c}_A + \frac{1}{2 c^2} \left( \frac{m_B}{M}\right)^2 ({\mathbf v} \cdot {\mathbf S}^{\rm c}_A) {\mathbf v} + O(c^{-4}).
\end{align}
\es
Since the spin variables differ at 1PN order, the equations of motion (but not the equations of precession) will have the same form for the 1.5PN and 2PN spin terms (aside from the replacements ${\mathbf S}^{\rm nc}_A \leftrightarrow {\mathbf S}^{\rm c}_A$), but the 2.5PN and higher-order spin terms will differ depending on the choice of spin variables. Throughout this paper the superscripts ``c'' and ``nc'' are sometimes dropped where either index would be appropriate.

The 2.5PN spin-orbit corrections to Eq.~\eqref{eq:PNspineqns} are given in Eq.~(5.7) of \cite{faye-buonanno-luc-higherorderspinI} in terms of the variables ${\mathbf S}^{\rm nc}$ and ${\mathbf \Sigma}^{\rm nc}$. The equivalent expressions in terms of the constant-magnitude spin variables are found by substituting the relations \eqref{eq:spintransform} into the 1.5PN SO term [Eq.~(5.7a) of \cite{faye-buonanno-luc-higherorderspinI}], and combining the result with the 2.5PN SO term in Eq.~(5.7b) of \cite{faye-buonanno-luc-higherorderspinI} (into which the substitutions ${\mathbf S} \rightarrow {\mathbf S}^{\rm c}$ and  ${\mathbf \Sigma} \rightarrow {\mathbf \Sigma}^{\rm c}$ can be made since we only require accuracy to relative 2.5PN order in the spin terms). The resulting SO contributions to Eq.~\eqref{eq:PNspineqns} in terms of the ``c'' spin variables are
\begin{widetext}
\begin{subequations}
\label{BSO}
\begin{equation}
\mathop{\mathbf{B}}_\text{SO}{}^{\!\!\mathrm{1.5PN}}
= \frac{1}{r^3} \bigg\{ \mathbf{n} \bigg[12 (S^{\rm c}, n, v) + 6 \frac{\delta m}{M} (\Sigma^{\rm c}, n, v) \bigg] + 9 (nv) \mathbf{n} \times \mathbf{S}^{\rm c} + 3 \frac{\delta m}{M} (nv) \mathbf{n} \times \mathbf{\Sigma}^{\rm c} - 7 \mathbf{v} \times \mathbf{S}^{\rm c} - 3 \frac{\delta m}{M} \mathbf{v} \times \mathbf{\Sigma}^{\rm c} \bigg\},
\end{equation}
\begin{multline}
\mathop{\mathbf{B}}_\text{SO}{}^{\!\!\mathrm{2.5PN}} = \frac{1}{r^3} \bigg\{ \mathbf{n} \bigg[(S^{\rm c}, n, v) \bigg( -30 \eta (nv)^2  + 24 \eta v^2 - \frac{M}{r} (44 + 25 \eta) \bigg) + \frac{\delta m}{M} (\Sigma^{\rm c}, n, v) \bigg( -15 \eta (nv)^2 + 12 \eta v^2 - \frac{M}{r} \left(24  + \frac{29}{2} \eta \right) \bigg) \bigg]
\\
 +  (nv) \mathbf{v} \bigg[ (S^{\rm c}, n, v) (-9 + 9 \eta) + \frac{\delta m}{M} (\Sigma^{\rm c}, n, v) (-3 + 6 \eta) \bigg] + \mathbf{n} \times \mathbf{v} \bigg[-\frac{3}{2}(nv) (vS^{\rm c}) (1 - \eta) - 8 \frac{M}{r} \eta (nS^{\rm c})
\\
 - \frac{\delta m}{M} \bigg(4 \frac{M}{r} \eta (n\Sigma^{\rm c}) + \frac{3}{2} (nv) (v\Sigma^{\rm c})  \bigg) \bigg]  + (nv) \mathbf{n} \times \mathbf{S}^{\rm c} \bigg[-\frac{45}{2} \eta (nv)^2 + 21 \eta v^2 - 7 \frac{M}{r} (4 + 3 \eta) \bigg]
\\
+ \frac{\delta m}{M} (nv) \mathbf{n} \times \mathbf{\Sigma}^{\rm c} \bigg[- 15 \eta (nv)^2  + 12 \eta v^2 - \frac{M}{r} \left( 12 + \frac{23}{2} \eta \right) \bigg] + \mathbf{v} \times \mathbf{S}^{\rm c}  \bigg[\frac{33}{2} \eta (nv)^2 + \frac{M}{r} (24 + 11 \eta) - 14 \eta v^2 \bigg]
\\
+ \frac{\delta m}{M} \mathbf{v} \times \mathbf{\Sigma}^{\rm c} \bigg[9 \eta (nv)^2 - 7 \eta v^2 + \frac{M}{r} (12 + \frac{11}{2} \eta) \bigg] \bigg\}.
\end{multline}
\end{subequations}
\end{widetext}
In the above equations we define additional notation following \cite{faye-buonanno-luc-higherorderspinI}: the unit vector ${\mathbf n}={\mathbf x}/r$ points in the direction of the relative separation vector ${\mathbf x}={\mathbf y}_1-{\mathbf y}_2$; ${\mathbf v}=\dot{{\mathbf x}}$ denotes the relative orbital velocity; scalar products of vectors are denoted by $(ab)\equiv {\mathbf a} \cdot {\mathbf b}$; and the mixed product of three vectors is denoted by $(a,b,c)\equiv{\mathbf a} \cdot ({\mathbf b} \times {\mathbf c})$.

The sum of the spin-spin and quadrupole-monopole terms is given in Eq.~(3.8) of \cite{racine-buonanno-kidder-spinningrecoil},
\be
\label{eq:SS-QMterm}
\mathop{\mathbf{B}}_\text{SS+QM}{}^{\!\!\!\!\!\!\!\!\mathrm{2PN}} = -\frac{3}{2M r^4} \left\{ \! \left[ ({\mathbf S}_0^{\rm c})^2 - 5 (nS_0^{\rm c})^2 \right] {\mathbf n} + 2(nS_0^{\rm c}) {\mathbf S}_0^{\rm c} \right\},
\ee
where
\be
{\mathbf S}_0^{\rm c} \equiv 2{\mathbf S}^{\rm c} + \frac{\delta m}{M} {\mathbf \Sigma}^{\rm c} = \left(1+ \frac{m_2}{m_1} \right) {\mathbf S}_1^{\rm c} + \left(1+ \frac{m_1}{m_2} \right) {\mathbf S}_2^{\rm c}.
\ee
Note that Eq.~\eqref{eq:SS-QMterm} has the same form in terms of the ``nc'' spin variables; it is also only valid for Kerr BHs as the value for the Kerr quadrupole moment was used.
Higher-order spin-spin corrections have recently been computed in Refs.~\cite{porto-rothstein-spinspinPRL2006,porto-rothstein-commentsteinhoff2007,porto-rothstein-spinspin3PN-PRD2008,*porto-rothstein-spinspin3PN-PRD2008-erratum,porto-rothstein-spin1spin1-PRD2008,*porto-rothstein-spin1spin1-PRD2008-erratum,steinhoff-hergt-schafer-PRD2008,hergt-schafer-PRD2008,steinhoff-schafer-PRDcomment2009,schaefer-confproc-massmotion2009,hergt-steinhoff-schafer-CQG2010,rothe-schafer-JMathPh2010,levi-spinspin-PRD2010}, but the explicit equations of motion have not yet been derived.
\subsection{\label{sec:nonprecess}Restriction to the nonprecessing case}
Now we restrict to nonprecessing orbits in which the individual spin vectors ${\mathbf S}_A$ are aligned or antialigned with the direction of the Newtonian orbital angular momentum vector ${\bm \ell} \equiv {\mathbf L}_{\rm N}/|{\mathbf L}_{\rm N}|$. We additionally define the unit vector ${\bm \lambda} \equiv {\bm \ell} \times {\mathbf n}$. Vectors can then be decomposed on the orthonormal basis $\{ {\mathbf n}, {\bm \lambda}, {\bm \ell} \}$ as in ${\mathbf S} = S_n {\mathbf n} + S_{\lambda} {\bm \lambda} + S_{\ell} {\bm \ell}$; similar relations hold for ${\mathbf \Sigma}$ and ${\mathbf S}_0$ (in either spin representation), as well as for ${\mathbf v}$. The restriction to nonprecessing orbits having a fixed orbital plane in the direction of ${\bm \ell}$ then implies the following relations:
\bs
\begin{align}
\label{eq:veq}
{\mathbf v} &= \dot{r} {\mathbf n} + r \dot{\varphi} {\bm \lambda}, \;\;\;\;\; v^2 = \dot{r}^2 + r^2 \dot{\varphi}^2, \\
(nv) &=\dot{r}, \;\;\;\;\; {\mathbf n} \times {\mathbf v} = r \dot{\varphi} {\bm \ell}, \\
{\mathbf S} &= S_{\ell} {\bm \ell}, \;\;\;\;\; {\mathbf \Sigma} = \Sigma_{\ell} {\bm \ell}, \;\;\;\;\; {\mathbf S}_0 = S_{0,\ell} {\bm \ell}, \\
(S, n, v) &= r \dot{\varphi} S_{\ell}, \;\;\;\;\; (\Sigma, n, v) = r \dot{\varphi} \Sigma_{\ell}, \\
{\mathbf n} \times {\mathbf S} &= -S_{\ell} {\bm \lambda}, \;\;\;\;\; {\mathbf n} \times {\mathbf \Sigma} = -\Sigma_{\ell} {\bm \lambda}, \\
{\mathbf v} \times {\mathbf S} &= S_{\ell} (r \dot{\varphi} {\mathbf n} - \dot{r} {\bm \lambda}),
\;\;\;\;\;
{\mathbf v} \times {\mathbf \Sigma} = \Sigma_{\ell} (r \dot{\varphi} {\mathbf n} - \dot{r} {\bm \lambda}), \\
(nS) &= (n\Sigma) = (vS) = (v\Sigma)=(nS_0)=0.
\end{align}
\es

The above relations allow the conservative PN two-body equations of motion to be put in the following form:
\be
\label{eq:PNeqnofmotion}
\frac{d{\mathbf v}}{dt} = -\frac{M}{r^2} \left[ (1+{\mathcal A}^{\rm tot}) {\mathbf n} + {\mathcal B}^{\rm tot} {\bm \lambda} \right],
\ee
where
\bs
\label{eq:AtotBtot}
\begin{align}
{\mathcal A}^{\rm tot} &= {\mathcal A}^{\rm NS} + {\mathcal A}^{\rm SO}_{\rm 1.5PN} + {\mathcal A}^{\rm SO}_{\rm 2.5PN} + {\mathcal A}^{\rm SS+QM}_{\rm 2PN} \\
{\mathcal B}^{\rm tot} &= {\mathcal B}^{\rm NS} + {\mathcal B}^{\rm SO}_{\rm 1.5PN} + {\mathcal B}^{\rm SO}_{\rm 2.5PN} + {\mathcal B}^{\rm SS+QM}_{\rm 2PN}.
\end{align}
\es
The nonspin terms ${\mathcal A}^{\rm NS}$ and ${\mathcal B}^{\rm NS}$ have been explicitly calculated by various authors. The results can be found in Eqs.~(181)-(196) of Blanchet's review article \cite{blanchetLRR}. Denoting Blanchet's expressions by ${\mathcal A}_{\rm B, NS}$ and ${\mathcal B}_{\rm B, NS}$, ignoring the dissipative terms at 2.5PN and 3.5PN orders, and using the form of the equations without the 3PN logarithmic terms, the nonspin terms in Eqs.~\eqref{eq:AtotBtot} are related to Blanchet's by
\bs
\begin{align}
{\mathcal A}^{\rm NS} &= {\mathcal A}_{\rm B, NS} + \dot{r} {\mathcal B}_{\rm B, NS} ,\\
{\mathcal B}^{\rm NS} &= r \dot{\varphi} {\mathcal B}_{\rm B, NS} .
\end{align}
\es
The 1.5PN spin-orbit terms are found to be
\bs
\begin{align}
{\mathcal A}^{\rm SO}_{\rm 1.5PN} &=-\frac{M}{r} (r \dot{\varphi}) \left[ 5 \frac{S_{\ell}^{\rm c}}{M^2} + 3 \frac{\delta m}{M}  \frac{\Sigma_{\ell}^{\rm c}}{M^2} \right], \\
{\mathcal B}^{\rm SO}_{\rm 1.5PN} &= 2 \frac{M}{r} \dot{r} \left( \frac{S_{\ell}^{\rm c}}{M^2} \right),
\end{align}
\es
and have the same form in terms of the ``nc'' variables. The 2.5PN spin-orbit terms in both spin variables are
\begin{widetext}
\bs
\begin{align}
{\mathcal A}^{{\rm SO,}\,{\rm c}}_{\rm 2.5PN} &= \! \frac{M}{r} (r \dot{\varphi}) \! \left\{ \! \left[ \frac{M}{r} (20+14\eta) + \! \left(9 - \frac{11}{2} \eta \right) \dot{r}^2 - 10 \eta (r \dot{\varphi})^2 \right] \! \frac{S_{\ell}^{\rm c}}{M^2} + \! \left[ \frac{M}{r} (12 + 9\eta) + (3-5\eta) \dot{r}^2 - 5\eta (r\dot{\varphi})^2 \right] \! \frac{\delta m}{M} \frac{\Sigma_{\ell}^{\rm c}}{M^2} \! \right\} \! ,\\
{\mathcal A}^{\rm SO, nc}_{\rm 2.5PN} &= \! \frac{M}{r} (r \dot{\varphi}) \! \left\{ \! \left[ \frac{M}{r} (17+16\eta) + \! \left(9 - \frac{11}{2} \eta \right) \dot{r}^2 - 10 \eta (r \dot{\varphi})^2 \right] \! \frac{S_{\ell}^{\rm nc}}{M^2} + \! \left[ \frac{M}{r} (9 + 10\eta) + (3-5\eta) \dot{r}^2 - 5\eta (r\dot{\varphi})^2 \right] \! \frac{\delta m}{M} \frac{\Sigma_{\ell}^{\rm nc}}{M^2} \! \right\} \! ,
\end{align}
\es
\bs
\begin{align}
{\mathcal B}^{{\rm SO,\,c}}_{\rm 2.5PN} &= \frac{M}{r} \dot{r} \left\{ \left[ - 2\frac{M}{r} (2+ 5\eta) + \eta \dot{r}^2 + (9-2\eta) (r \dot{\varphi})^2 \right] \frac{S_{\ell}^{\rm c}}{M^2} + \left[ -6 \frac{M}{r} \eta - \eta \dot{r}^2 + (3-\eta) (r \dot{\varphi})^2 \right] \frac{\delta m}{M} \frac{\Sigma_{\ell}^{\rm c}}{M^2} \right\}, \\
{\mathcal B}^{\rm SO, nc}_{\rm 2.5PN} &= \frac{M}{r} \dot{r} \left\{ \left[ - 2\frac{M}{r} (2+ 3\eta) + \eta \dot{r}^2 + (9-2\eta) (r \dot{\varphi})^2 \right] \frac{S_{\ell}^{\rm nc}}{M^2} + \left[ -4 \frac{M}{r} \eta - \eta \dot{r}^2 + (3-\eta) (r \dot{\varphi})^2 \right] \frac{\delta m}{M} \frac{\Sigma_{\ell}^{\rm nc}}{M^2} \right\}.
\end{align}
\es
\end{widetext}
Finally, the spin-spin + quadrupole-monopole pieces are
\bs
\begin{align}
{\mathcal A}^{\rm SS+QM}_{\rm 2PN} &= \frac{3}{2} \left( \frac{M}{r} \right)^2 \left( \frac{S_{0,\ell}^{\rm c}}{M^2} \right)^2, \\
{\mathcal B}^{\rm SS+QM}_{\rm 2PN} &= 0,
\end{align}
\es
and have the same form in terms of the ``nc'' variables.
\subsection{\label{sec:deriveISCO}Perturbing the equations of motion}
Having simplified the equations of motion, we now wish to study perturbations about the circular orbit solutions. We first reexpress the equations explicitly in terms of the polar coordinates $(r,\varphi)$ of the relative position vector. Differentiating the expression for the velocity vector in Eq.~\eqref{eq:veq} and using $\dot{\mathbf n} = \dot{\varphi} {\bm \lambda}$ and $\dot{\bm \lambda} = -\dot{\varphi} {\mathbf n}$, the components of Eq.~\eqref{eq:PNeqnofmotion} along ${\mathbf n}$ and ${\bm \lambda}$ are given by
\bs
\label{eq:PNeqns-polarcoords}
\begin{align}
\label{eq:rddot}
\ddot{r} &= -\frac{M}{r^2} (1 + {\mathcal A}^{\rm tot} ) + r \dot{\varphi}^2 , \\
\label{eq:phiddot}
\ddot{\varphi} &= - \frac{1}{r} \left( \frac{M}{r^2} {\mathcal B}^{\rm tot} + 2 \dot{r} \dot{\varphi} \right).
\end{align}
\es
This system can be reexpressed in first-order form by defining $u\equiv \dot{r}$ and $\omega \equiv \dot{\varphi}$, resulting in three first-order equations in the variables $(r,u,\omega)$.

Circular orbits correspond to the conditions $\dot{r} = \dot{u} = \dot{\omega} =0$. In particular, the condition $\dot{u}=0$ and Eq.~\eqref{eq:rddot} imply the following implicit relationship for the circular orbital frequency:
\be
\label{eq:omega-cond}
\omega_0^2 = \frac{M}{r_0^3} [1+ {\mathcal A}_0^{\rm tot}(r_0,\omega_0)],
\ee
or, in terms of the PN parameter $x\equiv (M \omega_0)^{2/3}$,
\be
\label{eq:x0-implicit}
x = \gamma [1+ {\mathcal A}_0^{\rm tot}(\gamma,x)]^{1/3},
\ee
where a subscript $0$ refers to quantities evaluated along a circular orbit and we have defined another PN expansion parameter $\gamma\equiv M/r$.

Equation \eqref{eq:x0-implicit} provides an implicit relationship between the two PN expansion parameters $\gamma$ and $x$. Later, we shall need an explicit PN expansion for $\gamma$ in terms of $x$. To derive this relationship from \eqref{eq:x0-implicit}, we first substitute a 3PN series expansion with undetermined coefficients,
\be
\label{eq:gamma-expand}
\gamma = x ( 1 + c_{1} x + c_{1.5} x^{3/2} + c_{2} x^{2} + c_{2.5} x^{5/2} + c_{3} x^{3} ),
\ee
into the right-hand-side of Eq.~\eqref{eq:x0-implicit}. Next we series expand the result in $x$ to 3PN order, and equate the coefficients of like powers of $x$ on both sides of the equation. This results in a linear system of 5 equations for the 5 unknowns in Eq.~\eqref{eq:gamma-expand}. Solving this system easily yields
\begin{multline}
\label{eq:gamma-x-spin}
\gamma = x \left\{ 1 + x \left( 1 - \frac{\eta}{3} \right) + x^{3/2} \left( \frac{5}{3} \frac{S_{\ell}^{\rm c}}{M^2} + \frac{\delta m}{M} \frac{\Sigma_{\ell}^{\rm c}}{M^2} \right)
\right. \\
+ x^2 \left[ 1 - \frac{65}{12} \eta - \frac{1}{2} \left( \frac{S_{0,\ell}^{\rm c}}{M^2} \right)^2 \right]
\\
+ x^{5/2} \left[ \left( \frac{10}{3} +\frac{8}{9} \eta \right) \frac{S_{\ell}^{\rm c}}{M^2} + 2 \frac{\delta m}{M} \frac{\Sigma_{\ell}^{\rm c}}{M^2} \right]
\\ \left.
+ x^3 \left[ 1 + \left( -\frac{2203}{2520} - \frac{41}{192}\pi^2 \right) \eta + \frac{229}{36} \eta^2 +\frac{\eta^3}{81} \right] \right\}.
\end{multline}
In terms of the nonconstant spin-magnitude variables, the 2.5PN order term in the above equation should be replaced with [see Eq.~(6.3) of \cite{faye-buonanno-luc-higherorderspinII,*faye-buonanno-luc-higherorderspinIIerratum,*faye-buonanno-luc-higherorderspinIIerratum2}]
\be
\label{eq:gamma-x-spin-nc}
+ x^{5/2} \left[ \left( \frac{13}{3} +\frac{2}{9} \eta \right) \frac{S_{\ell}^{\rm nc}}{M^2} + \left( 3-\frac{\eta}{3} \right) \frac{\delta m}{M} \frac{\Sigma_{\ell}^{\rm nc}}{M^2} \right],
\ee
while the 1.5PN and 2PN order spin terms have the same form with ``c'' replaced by ``nc''.

Now we examine linear perturbations to the equations of motion \eqref{eq:PNeqns-polarcoords} about circular orbits parametrized by $(r_0, \omega_0)$. Introducing a small expansion parameter $\epsilon$, we substitute the following expansions into Eqs.~\eqref{eq:PNeqns-polarcoords}:
\bs
\begin{align}
r &= r_0 + \epsilon \delta r, \\
u &= 0 + \epsilon \delta u, \\
\omega &= \omega_0 + \epsilon \delta \omega,
\end{align}
\es
and linearize. In doing so we expand ${\mathcal A}^{\rm tot}$ as
\be
{\mathcal A}^{\rm tot} = {\mathcal A}^{\rm tot}_0 + \epsilon \frac{\partial {\mathcal A}^{\rm tot}}{\partial r}\bigg|_{0} \delta r  + \epsilon \frac{\partial {\mathcal A}^{\rm tot}}{\partial u}\bigg|_{0} \delta u + \epsilon \frac{\partial {\mathcal A}^{\rm tot}}{\partial \omega}\bigg|_{0} \delta \omega,
\ee
and likewise for ${\mathcal B}^{\rm tot}$. From the explicit form of ${\mathcal A}^{\rm tot}$ and ${\mathcal B}^{\rm tot}$, one can verify that
\be
\frac{\partial {\mathcal A}^{\rm tot}}{\partial u}\bigg|_0 = \frac{\partial {\mathcal B}^{\rm tot}}{\partial r}\bigg|_0 = \frac{\partial {\mathcal B}^{\rm tot}}{\partial \omega}\bigg|_0 = 0.
\ee
Then, at $O(\epsilon^0)$, the equations of motion reduce to Eq.~\eqref{eq:omega-cond} and ${\mathcal B}^{\rm tot}_0 =0$. At $O(\epsilon^1)$, we have the system
\bs
\label{eq:linearpert}
\begin{align}
\dot{\delta r} &= \delta u, \\
\dot{\delta u} &= \alpha_0 \delta r + \beta_0 \delta \omega, \\
\dot{\delta \omega} &= \gamma_0 \delta u, \;\;\;\;\; \text{with}
\end{align}
\es
\bs
\label{eq:alpha-beta-gamma}
\begin{align}
\alpha_0 &= 3\omega_0^2 -\frac{M}{r_0^2} \frac{\partial {\mathcal A}^{\rm tot}}{\partial r} \bigg|_0, \\
\beta_0 &= 2 r_0 \omega_0 -\frac{M}{r_0^2} \frac{\partial {\mathcal A}^{\rm tot}}{\partial \omega} \bigg|_0, \\
\gamma_0 &= -\frac{1}{r_0} \left( 2\omega_0 + \frac{M}{r_0^2} \frac{\partial {\mathcal B}^{\rm tot}}{\partial u} \bigg|_0 \right),
\end{align}
\es
where $\gamma_0$ is not related to the $\gamma\equiv M/r$ defined earlier.

Now we assume a perturbation of the form $\delta q = E_q e^{i \lambda t}$ [where $q=(r,u,\omega)$] and substitute into Eqs.~\eqref{eq:linearpert}, resulting in a linear algebraic system for the $E_q$ and the eigenvalue $\lambda$. A trivial solution corresponding to $\lambda=0$ is $E_u=0$ and $E_r = -(\beta_0/\alpha_0) E_{\omega}$; this represents a nonoscillatory displacement from one circular orbit to another. The remaining eigenvalues are $\lambda = \pm [-(\alpha_0 + \beta_0 \gamma_0)]^{1/2}$. If the argument of the square-root is positive, then the resulting solutions are stable. The condition for the existence of stable circular orbits can therefore be expressed as
\be
\label{eq:C0cond}
C_0 \equiv -\alpha_0 - \beta_0 \gamma_0 > 0,
\ee
and the equality $C_0 =0$ defines the ISCO.

Using Eqs.~\eqref{eq:alpha-beta-gamma} and \eqref{eq:AtotBtot}, eliminating $r$ via \eqref{eq:gamma-x-spin}, and expanding to the appropriate PN order, one can express the stability condition explicitly in terms of $x$, yielding the following gauge-invariant condition for the ISCO:
\begin{multline}
\label{eq:C0hat-spin}
\hat{C}_0 \equiv \frac{M^2}{x^3} C_0 = 1- 6x + x^{3/2} \left( 14 \frac{S_{\ell}^{\rm c}}{M^2} + 6 \frac{\delta m}{M} \frac{\Sigma_{\ell}^{\rm c}}{M^2} \right)
\\
+ x^2 \left[ 14\eta - 3 \left( \frac{S_{0,\ell}^{\rm c}}{M^2} \right)^2 \right]
\\
+ x^{5/2} \left[ - \frac{S_{\ell}^{\rm c}}{M^2} (22+32\eta) - \frac{\delta m}{M} \frac{\Sigma_{\ell}^{\rm c}}{M^2} (18+15\eta) \right]
\\
+ x^3 \left[ \left( \frac{397}{2} - \frac{123}{16}\pi^2  \right) \eta - 14 \eta^2 \right].
\end{multline}
In terms of the nonconstant-magnitude spin variables, the 2.5PN spin-orbit term is replaced with
\be
\label{eq:C0hat-spin-nc}
+ x^{5/2} \left[ - \frac{S_{\ell}^{\rm nc}}{M^2} (13+30\eta) - \frac{\delta m}{M} \frac{\Sigma_{\ell}^{\rm nc}}{M^2} (9+14\eta) \right],
\ee
while the 1.5PN and 2PN spin terms have the same form with ``c'' relabeled to ``nc''.
Note that in the nonspinning case, Eq.~\eqref{eq:C0hat-spin} reduces to the 3PN gauge-invariant stability condition of Blanchet and Iyer \cite{blanchetiyer3PN} [their Eq.~(6.41) or Eq.~\eqref{eq:C0invar} here].
\section{\label{sec:kerriscocompare}Comparison with the Kerr ISCO}
In the nonspinning case, Eq.~\eqref{eq:C0hat-spin} reduces in the test-mass limit to
\be
\hat{C}_0 = 1-6x.
\ee
The PN ISCO criterion $\hat{C}_0=0$ in this case clearly reproduces the exact Schwarzschild ISCO, $x=1/6$. We wish to determine if Eq.~\eqref{eq:C0hat-spin} similarly reproduces the Kerr ISCO.

Recall that the Kerr ISCO radius in Boyer-Lindquist coordinates is given by \cite{bpt}
\begin{align}
\label{eq:risco-kerr}
\frac{r_{\rm isco}^{\rm K}}{m_2} &= 3+Z_2 -\sign(\chi_2^{\rm K}) [(3-Z_1)(3+Z_1+2Z_2)]^{1/2}, \nonumber \\
Z_1 &= 1+ [1-(\chi_2^{\rm K})^2]^{1/3}[(1+\chi_2^{\rm K})^{1/3} + (1-\chi_2^{\rm K})^{1/3}], \nonumber \\
Z_2 &= [3(\chi_2^{\rm K})^2 + Z_1^2]^{1/2},
\end{align}
where the mass of the Kerr BH is denoted $m_2$, and its dimensionless spin is $\chi_2^{\rm K} \in [-1,1]$ (with negative values corresponding to point-particles with retrograde orbital motion).\footnote{In the notation of the previous section, the BH spin angular momentum is ${\mathbf S}^{\rm K}_2 \equiv \chi_2^{\rm K} m_2^2 {\hat{\mathbf s}}^{\rm K}_2$, where, in our restriction to nonprecessing circular orbits, the orbital angular momentum points in the ${\bm \ell} = \hat{\mathbf z}$ direction and we choose ${\hat{\mathbf s}}^{\rm K}_2=\hat{\mathbf z}$.}
An expression equivalent to Eq.~\eqref{eq:risco-kerr} can be found by differentiating the reduced particle energy \cite{bpt}
\be
\label{eq:Ekerr}
\tilde{E} \equiv \frac{E}{m_1} = \frac{1-2w_{\rm BL}+\chi_2^{\rm K} w_{\rm BL}^{3/2}}{\sqrt{1-3w_{\rm BL}+2\chi^{\rm K}_2 w_{\rm BL}^{3/2}}},
\ee
where $w_{\rm BL} \equiv m_2/r_{\rm BL}$ and $r_{\rm BL}$ is the Boyer-Lindquist radial coordinate. Some simple algebraic manipulation of $d\tilde{E}/dr_{\rm BL}=0$ yields
\be
\label{eq:C0kerr-rBL}
\hat{C}_0^{\rm K} \equiv 1 - 6w_{\rm BL} + 8 \chi_2^{\rm K} w_{\rm BL}^{3/2} - 3 (\chi_2^{\rm K})^2 w_{\rm BL}^2 =0.
\ee
Solving this equation for $r_{\rm BL}$ produces results identical to Eq.~\eqref{eq:risco-kerr}.
But note that since $w_{\rm BL}$ depends on a coordinate radius, Eq.~\eqref{eq:C0kerr-rBL} is clearly not a gauge-invariant expression.

To derive a gauge-invariant version of Eq.~\eqref{eq:C0kerr-rBL}, we first define the variable $X\equiv |m_2 \Omega^{\rm K}|^{2/3}$, which is analogous to the PN parameter $x$ (in the test-mass limit, $x\rightarrow X$). The frequency $\Omega^{\rm K} \equiv d\varphi/dt$ refers to the circular-orbit angular frequency seen by a distant observer and follows from the Kerr geodesic equations [Eq.~(2.16) of \cite{bpt}]:
\be
\label{eq:Omega-kerr}
m_2 \Omega^{\rm K} = \sign(\chi_2) \frac{w_{\rm BL}^{3/2}}{1+\chi_2^{\rm K} w_{\rm BL}^{3/2}}.
\ee
Defining $\beta \equiv 1-\chi_2^{\rm K} X^{3/2}$, we invert Eq.~\eqref{eq:Omega-kerr} to obtain
\be
\label{eq:wBL}
w_{\rm BL}=\frac{X}{\beta^{2/3}}.
\ee
Substituting this result into Eq.~\eqref{eq:C0kerr-rBL}, we arrive at the gauge-invariant relation
\be
\label{eq:C0kerr}
\hat{C}_0^{\rm K} \equiv 1 - \frac{X}{\beta^{2/3}} \left[ 6 - \chi_2^{\rm K} \frac{X^{1/2}}{\beta^{1/3}} \left( 8 - 3\chi_2^{\rm K} \frac{X^{1/2}}{\beta^{1/3}} \right) \right].
\ee
For $\chi_2^{\rm K}=0$ we easily obtain the Schwarzschild value for the ISCO frequency ($X=1/6$). One can verify numerically that solving $\hat{C}_0^{\rm K}=0$ as a function of $\chi_2^{\rm K}$ reproduces the ISCO frequency computed from Eqs.~\eqref{eq:risco-kerr} and \eqref{eq:Omega-kerr} for all values of $\chi_2^{\rm K} \in [-1,1]$.

Now we wish to compare the test-mass limit of the ISCO condition derived in Eq.~\eqref{eq:C0hat-spin} with the gauge-invariant Kerr ISCO expression in Eq.~\eqref{eq:C0kerr}. Note that Eq.~\eqref{eq:C0kerr} is valid for arbitrary spin, while Eq.~\eqref{eq:C0hat-spin} is limited by the PN order to which spin terms have been computed in the equations of motion (currently 2.5PN order). To allow a meaningful comparison, we must expand Eq.~\eqref{eq:C0kerr} in the spin parameter $\chi_2^{\rm K}$, yielding
\begin{multline}
\label{eq:C0kerr-smallspin}
\hat{C}_0^{\rm K} = 1 - 6 X + \chi_2^{\rm K} ( 8 X^{3/2} - 4X^{5/2} )
\\
+ (\chi_2^{\rm K})^2 \left(- 3X^2 + 8X^3 -10 X^4/3 \right) + O[(\chi_2^{\rm K})^3].
\end{multline}
We can also perform a PN expansion of Eq.~\eqref{eq:C0kerr} in $X$, which results in
\begin{multline}
\label{eq:C0kere-smallX}
\hat{C}_0^{\rm K} = 1 - 6X + 8\chi_2^{\rm K} X^{3/2} - 3(\chi_2^{\rm K})^2 X^2
\\
- 4\chi_2^{\rm K} X^{5/2} + 8(\chi_2^{\rm K})^2 X^3 +O[(\chi_2^{\rm K})^3 X^{7/2}].
\end{multline}
Note that both expansions give consistent results at the appropriate orders in $\chi_2^{\rm K}$ and $X$. This is especially interesting because in Eq.~\eqref{eq:C0kerr-smallspin}, no PN expansion has been made. It also suggests the presence of additional self-spin terms at 3PN and 4PN orders in the equations of motion (in addition to the currently known 2PN-order terms). Equation \eqref{eq:C0kere-smallX} suggests that cubic self-spin interaction terms will not appear until 3.5PN order.

The above expansions can now be compared with the test-mass limit of Eqs.~\eqref{eq:C0hat-spin} and \eqref{eq:C0hat-spin-nc}. Taking $\eta \rightarrow 0$, $\delta m/M \rightarrow -1$, and $(S_{\ell}/M^2,\Sigma_{\ell}/M^2, S_{0,\ell}/M^2) \rightarrow \chi_2$, the result is
\be
\label{eq:C0spin1}
\hat{C}_0 = 1 - 6 x + 8 \chi_2 x^{3/2} - 3 \chi_2^2 x^2 - 4 \chi_2 x^{5/2} + O(x^3).
\ee
This is valid for either choice of spin variable ($\chi_2^{\rm nc}$ or $\chi_2^{\rm c}$).
Comparing with Eq.~\eqref{eq:C0kerr-smallspin} (and identifying $X$ with $x$ and $\chi_2^{\rm K}$ with $\chi_2$), we see that the PN gauge-invariant ISCO condition \eqref{eq:C0spin1} agrees with the Kerr ISCO condition up to the PN order (2.5PN) to which we know the spin terms in the PN equations of motion.

Figure \ref{fig:isco-compare} compares different methods for computing the ISCO frequency (in the test-mass limit): (i) the exact Kerr expression [computed from solving Eq.~\eqref{eq:C0kerr} or plugging Eq.~\eqref{eq:risco-kerr} into Eq.~\eqref{eq:Omega-kerr}]; (ii) solving the gauge-invariant ISCO condition in Eq.~\eqref{eq:C0hat-spin} or \eqref{eq:C0spin1}; and (iii) finding the minimum of the PN circular-orbit energy with nonspin terms to 3PN order and spin terms to 2.5PN order [Eq.~\eqref{eq:E3PN}]. [The ISCO frequencies for approaches (i) and (ii) are also listed in Table \ref{tab:isco-quantities}.] The method using the gauge-invariant condition $\hat{C}_0$ agrees exceptionally well for all spins up to $\chi_2 \lesssim 0.5$. In the nonspinning case ($\chi_2=0$), the agreement is exact. For nonzero spins, agreement with the exact Kerr result is limited by the fact that we only know the spin terms in the equations of motion to 2.5PN order. Note also that for small $|\chi_2|$, the error is symmetric about $\chi_2=0$. This is in contrast with the ISCO computed from the 3PN energy function, for which the error increases (nearly) monotonically with increasing ISCO frequency (or decreasing radius). This indicates that the ISCO computed via $\hat{C}_0$ is limited not by finite-PN corrections but by finite-spin corrections.

In Appendix \ref{app:spincheck} we examine how other PN expressions agree with their Kerr-spacetime counterparts. We find that test-mass limits of the circular-orbit energy and the Keplerian relation $\gamma(x)$ agree with their Kerr analogs if we identify $\chi_2^{\rm K}$ with either choice of spin variable. However, the PN orbital angular momentum only agrees with its Kerr analog if we identify $\chi_2^{\rm K}$ with $\chi_2^{\rm c}$.
\section{\label{sec:ISCOshift}Conservative shifts in the ISCO}
Consider the general behavior of the ISCO frequency when the test-particle has a non-negligible mass and spin (but assume that all spins are aligned or antialigned with the orbital angular momentum). The ISCO frequency can be split into the following pieces:\footnote{In the remainder of this paper and unless stated otherwise, all of the spin variables refer to the constant-magnitude spins.}
\begin{multline}
\label{eq:ISCO-expand-schematic}
m_2 \Omega = \hat{\Omega}^{\rm K}(\chi_2) + \delta\hat{\Omega}^{\rm GSF}(\chi_2,q)
\\
+ \delta \hat{\Omega}^{\rm COspin}(\chi_2,q,\chi_1) + \delta \hat{\Omega}^{\rm GSF+COspin}(\chi_2,q,\chi_1),
\end{multline}
where $\hat{\Omega}^{\rm K}$ is the Kerr ISCO frequency in units of $m_2$ [given by Eqs.~\eqref{eq:risco-kerr} and \eqref{eq:Omega-kerr}, or Eq.~\eqref{eq:C0kerr}], $\delta \hat{\Omega}^{\rm GSF}$ and $\delta \hat{\Omega}^{\rm COspin}$ are corrections to this frequency (also in units of $m_2$) due to the conservative GSF and the spin of the smaller compact object, and $\delta \hat{\Omega}^{\rm GSF+COspin}$ is a correction that results from cross-terms between both effects. If we assume that the mass ratio $q\equiv m_1/m_2\leq 1$ is small, then we can rewrite Eq.~\eqref{eq:ISCO-expand-schematic} as
\begin{multline}
\label{eq:ISCO-expand-schematic2}
m_2 \Omega = \hat{\Omega}^{\rm K}(\chi_2) [ 1+ q c'^{\rm GSF}(\chi_2) + q \chi_1 c^{\rm COspin}(\chi_2)
\\
+ O(q^2) + O(\chi_1 q^2) + O(\chi_1^2 q^2) ].
\end{multline}
Multiplying by $M/m_2$ and using $\eta=q + O(q^2)$ \cite{damour-GSF} yields
\begin{multline}
\label{eq:ISCO-expand-schematic3}
\tilde{\Omega} \equiv M\Omega = \hat{\Omega}^{\rm K}(\chi_2) [ 1+ \eta c^{\rm GSF}(\chi_2) + \eta \chi_1 c^{\rm COspin}(\chi_2)
\\
+ O(\eta^2) + O(\chi_1 \eta^2) +O(\chi_1^2 \eta^2)],
\end{multline}
where $c^{\rm GSF}=1+c'^{\rm GSF}$ was labeled $c_{\Omega}^{\rm ren}$ in \cite{damour-GSF,favata-iscostudy} for the $\chi_2=0$ case.

In the remainder of this section, we shall concern ourselves with the calculation of the coefficients $c^{\rm GSF}(\chi_2)$ and $c^{\rm COspin}(\chi_2)$ via the improved spinning-EOB Hamiltonian of \cite{barausse-buonanno-spinEOB} and the new gauge-invariant PN ISCO condition in Eqs.~\eqref{eq:C0hat-spin}. In particular, we note that the improved EOB Hamiltonian is constructed such that the coefficients $c^{\rm GSF}_{\rm EOB}(0)$ and $c^{\rm COspin}_{\rm EOB}(\chi_2)$ are \emph{exact}.
\subsection{\label{sec:spinEOB}The improved effective-one-body Hamiltonian for spinning binaries}
Recently, Barausse and Buonanno \cite{barausse-buonanno-spinEOB} have constructed a new EOB Hamiltonian with the following features: (i) In the test-particle limit, the Hamiltonian reduces to the exact Hamiltonian of a spinning test-body in the Kerr spacetime \cite{barausse-racine-buonanno-PRD2009} (to linear order in the test-particle's spin; this limit of the EOB Hamiltonian produces equations of motion and precession that are equivalent to the Papapetrou-Mathisson-Dixon equations \cite{papapetrou-spinningI-1951,papapetrou-spinningII-1951,mathisson-1931-original,*mathisson-1931-reprint,mathisson-1937-original,*mathisson-1937-reprint,dixon-extendedbodiesI-PRSocA1970,dixon-extendedbodiesII-PRSocA1970,dixon-extendedbodiesIII-PRSocA1974}). (ii) When PN-expanded, the EOB Hamiltonian reproduces the 2PN spin-spin and 1.5PN and 2.5PN spin-orbit couplings for arbitrary mass ratios. (iii) The Hamiltonian includes an adjustable function $K(\eta)$ that appears in the spinning generalization of the effective metric function $A(r)$ [see Eqs.~(6.9)--(6.11) of \cite{barausse-buonanno-spinEOB}]; this function is adjusted to enforce agreement with the Barack-Sago conservative GSF shift in the Schwarzschild ISCO \cite{barack-sago_isco,barack-sago-eccentricselfforce}. [But note that this adjustment does not guarantee good agreement with the (yet uncalculated) conservative GSF shift in the Kerr ISCO.] (iv) For arbitrary mass ratios, this improved EOB Hamiltonian provides a well-defined prescription to compute the conservative two-body dynamics and spin precession. (v) Finally, in the case of aligned or antialigned spins, this conservative dynamics produces a well-behaved ISCO for any mass ratio.

The improved EOB Hamiltonian of \cite{barausse-buonanno-spinEOB} is complicated to write out explicitly. For the case of equatorial (nonprecessing) orbits with spins aligned or antialigned with the orbital angular momentum, one can construct the Hamiltonian by starting with Eq.~(6.1) of \cite{barausse-buonanno-spinEOB} and carefully following their paper for the subsequent chain of definitions (see Appendix C of \cite{yunes-etal-EMRI-EOB-spin} for an alternate presentation). Once the EOB Hamiltonian is constructed, the ISCO angular frequency can be computed from Eqs.~(6.6)--(6.8) of \cite{barausse-buonanno-spinEOB}. Choosing units in which the total mass $M=1$, I constructed a numerical code which computes the ISCO frequency $\tilde{\Omega}_{\rm EOB}(\eta,\chi_1,\chi_2)$ given the reduced mass-ratio $\eta$, the spin of the test-particle $\chi_1$, and the BH spin $\chi_2$. By construction, the resulting EOB ISCO has three important properties: (i) in the test-particle limit it reduces to the Kerr ISCO [$\tilde{\Omega}_{\rm EOB}(0,0,\chi_2) = \hat{\Omega}^{\rm K}(\chi_2)$]; (ii) in the nonspinning case it reproduces the exact conservative GSF ISCO shift [$c_{\rm EOB}^{\rm GSF}(0)=c_{\Omega}^{\rm ren}\approx 1.251$]; and (iii) it correctly accounts for the conservative ISCO shift due to the test-particle's spin (this was explicitly verified in Appendix \ref{app:papa} by directly analyzing the Papapetrou equations).
\begin{table*}[t]
\caption{\label{tab:isco-quantities}ISCO quantities as a function of the dimensionless BH spin parameter $\chi_2$. The second column denotes the standard Kerr ISCO angular frequency in units of $m_2$ [Eqs.~\eqref{eq:risco-kerr} and \eqref{eq:Omega-kerr}]. The third column is the test-particle limit of the ISCO frequency computed from the gauge-invariant ISCO condition $\hat{C}_0$ [Eq.~\eqref{eq:C0hat-spin} or \eqref{eq:C0spin1}]. The fourth column is the conservative self-force ISCO shift parameter computed from the EOB ISCO frequency [Eq.~\eqref{eq:cGSF-EOB}]. The fifth column is the analogous quantity computed from the $\hat{C}_0$ ISCO condition [Eq.~\eqref{eq:cGSF-C0}]. The sixth column computes the ISCO shift parameter due to the spin of the test-particle (computed via the spinning-EOB ISCO frequency [Eq.~\eqref{eq:cCOspinEOB}], or directly from the Papapetrou equations [Appendix \ref{app:papa}]). The seventh column is the analogous quantity computed via the $\hat{C}_0$ ISCO condition [Eq.~\eqref{eq:cCOspinC0}]. Note the perfect agreement of several of these quantities in the $\chi_2=0$ case, and the closeness in their values for small $\chi_2 \lesssim 0.6$ (see also Figs.~\ref{fig:isco-compare} and \ref{fig:cEOB-PN}).}
\begin{ruledtabular}
\begin{tabular}{rrrrrrr}
$\chi_2$ & $\Omega^{\rm Kerr}_{\rm isco}$  & $\Omega_{C_0}^{\rm isco}$ & $c_{\rm EOB}^{\rm GSF}$ & $c_{C_0}^{\rm GSF}$ & $c_{\rm EOB}^{\rm COspin}$ & $c_{\rm PN}^{\rm COspin}$ \\
\hline
-0.99 &	0.038\,635 &	0.038\,015 &	0.9486 &	1.1903 &	0.2313 &	0.1945 \\
-0.9 &	0.040\,261 &	0.039\,681 &	0.9449 &	1.1961 &	0.2364 &	0.2020 \\
-0.8 &	0.042\,223 &	0.041\,694 &	0.9423 &	1.2043 &	0.2424 &	0.2110 \\
-0.7 &	0.044\,372 &	0.043\,901 &	0.9422 &	1.2148 &	0.2487 &	0.2205 \\
-0.6 &	0.046\,736 &	0.046\,331 &	0.9458 &	1.2282 &	0.2553 &	0.2308 \\
-0.5 &	0.049\,348 &	0.049\,016 &	0.9550 &	1.2453 &	0.2625 &	0.2417 \\
-0.4 &	0.052\,251 &	0.051\,998 &	0.9726 &	1.2670 &	0.2700 &	0.2534 \\
-0.3 &	0.055\,496 &	0.055\,325 &	1.0027 &	1.2948 &	0.2782 &	0.2657 \\
-0.2 &	0.059\,149 &	0.059\,057 &	1.0517 &	1.3303 &	0.2868 &	0.2788 \\
-0.1 &	0.063\,295 &	0.063\,266 &	1.1295 &	1.3759 &	0.2962 &	0.2923 \\
0.0 &	0.068\,041 &	0.068\,041 &	1.2513 &	1.4349 &	0.3062 &	0.3062 \\
0.1 &	0.073\,536 &	0.073\,492 &	1.4418 &	1.5116 &	0.3170 &	0.3199 \\
0.2 &	0.079\,979 &	0.079\,750 &	1.7400 &	1.6118 &	0.3287 &	0.3328 \\
0.3 &	0.087\,652 &	0.086\,978 &	2.2072 &	1.7434 &	0.3414 &	0.3435 \\
0.4 &	0.096\,974 &	0.095\,365 &	2.9338 &	1.9167 &	0.3551 &	0.3502 \\
0.5 &	0.108\,588 &	0.105\,125 &	4.0204 &	2.1441 &	0.3699 &	0.3499 \\
0.6 &	0.123\,568 &	0.116\,470 &	5.4310 &	2.4388 &	0.3856 &	0.3382 \\
0.7 &	0.143\,879 &	0.129\,564 &	6.3967 &	2.8098 &	0.4014 &	0.3097 \\
0.8 & 	0.173\,747 &	0.144\,421 &	4.2785 &	3.2524 &	0.4150 &	0.2590 \\
0.9 &	0.225\,442 &	0.160\,767 &	-3.3671 &	3.7337 &	0.4152 &	0.1837 \\
0.99 &	0.364\,410 &	0.176\,197 &	-23.763 &	4.1440 &	0.2937 &	0.0983 \\
\end{tabular}
\end{ruledtabular}
\end{table*}
\begin{figure*}[t]
$
\begin{array}{cc}
\includegraphics[angle=0, width=0.48\textwidth]{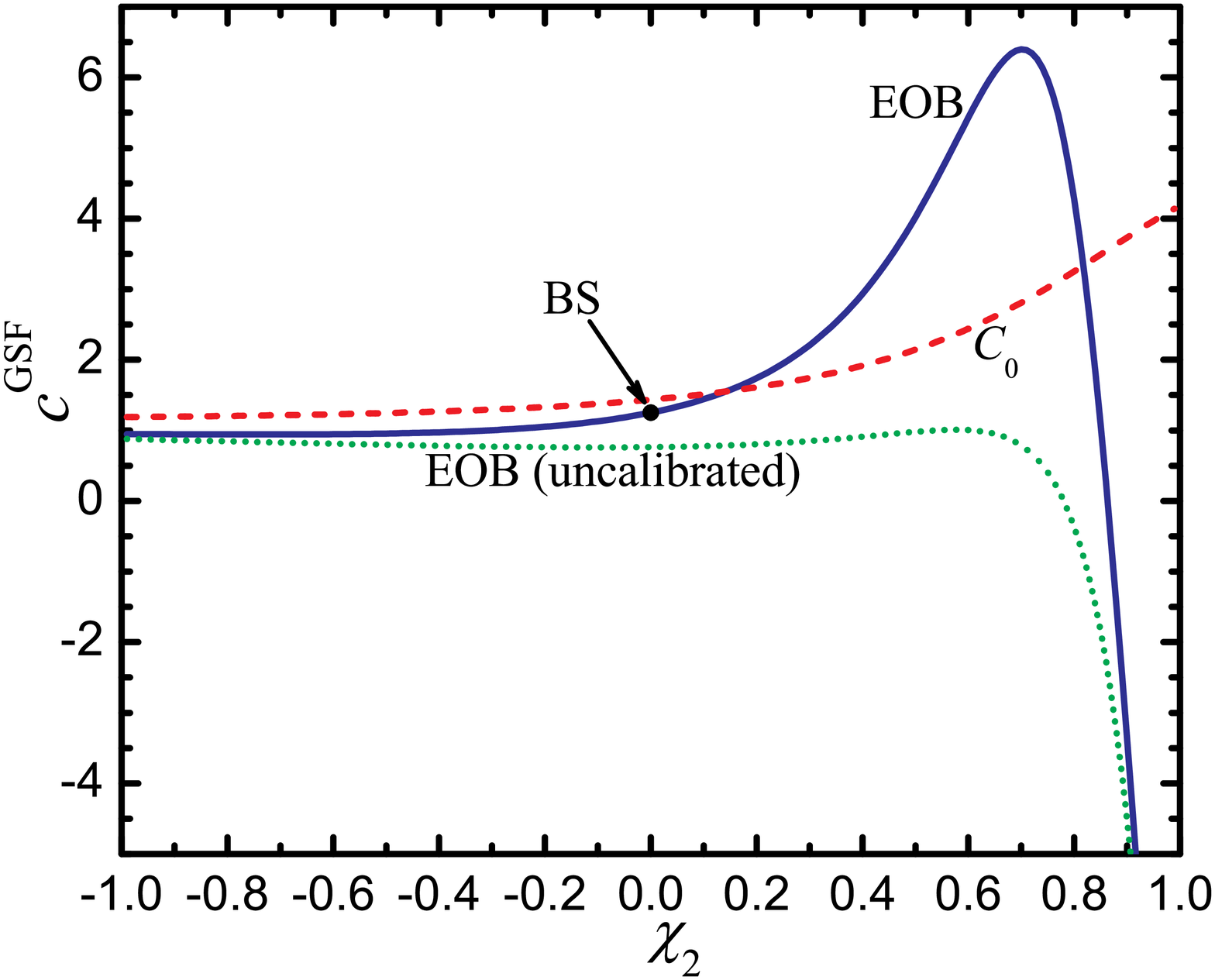} &
\includegraphics[angle=0, width=0.48\textwidth]{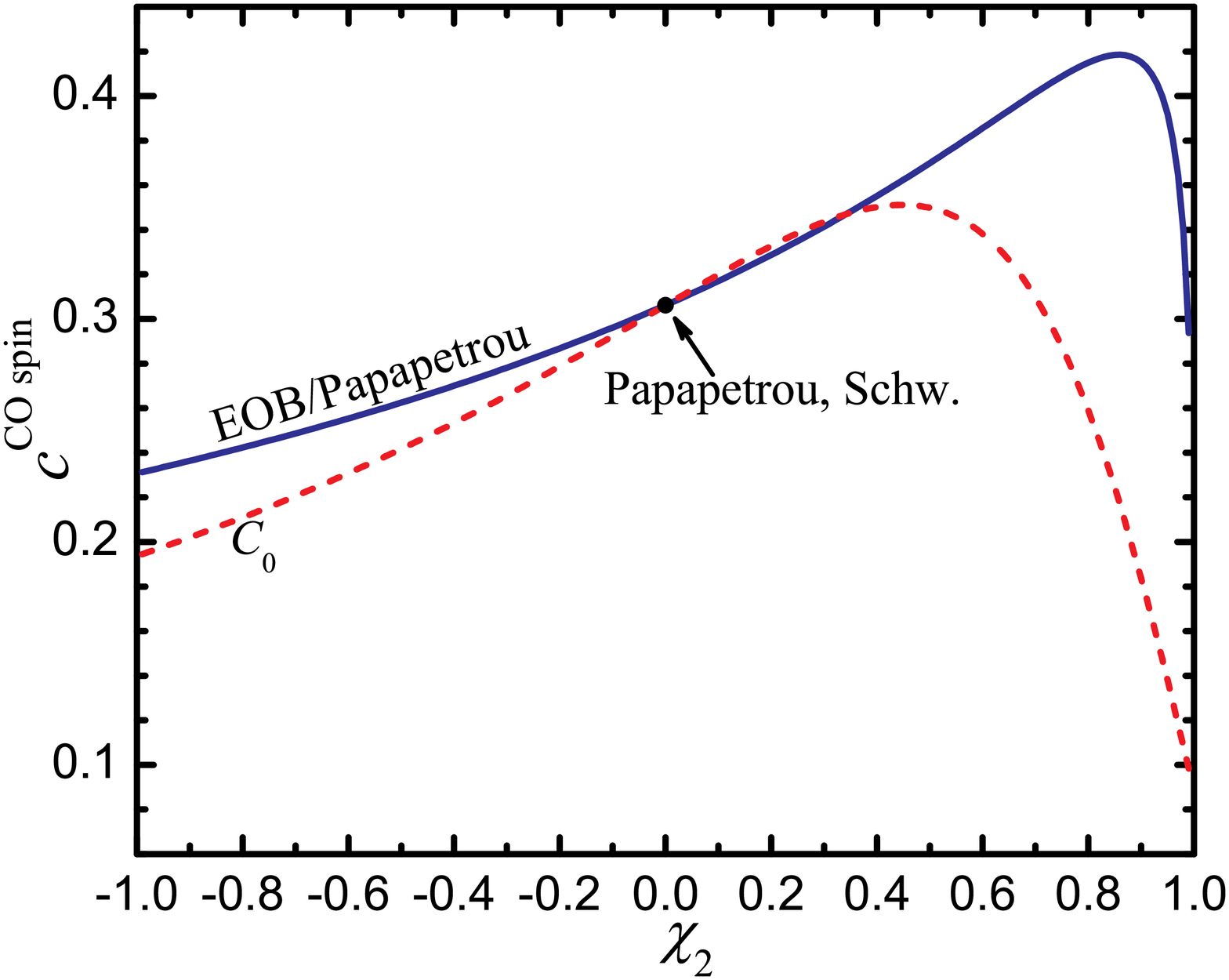}
\end{array}
$
\caption{\label{fig:cEOB-PN}(color online). ISCO shift parameters computed via the improved spinning-EOB Hamiltonian of \cite{barausse-buonanno-spinEOB} and the gauge-invariant ISCO condition in Eq.~\eqref{eq:C0hat-spin}. The left plot shows the ISCO shift due to the conservative gravitational self-force (GSF) as a function of the big BH spin $\chi_2$ (the test-particle is assumed to be nonspinning in this case). The solid (blue) ``EOB'' curve uses the Hamiltonian from \cite{barausse-buonanno-spinEOB} [which is fit to the exact Barack-Sago (BS) result in the nonspinning case] and Eq.~\eqref{eq:cGSF-EOB}. The dotted (green) ``EOB (uncalibrated)'' curve also uses this Hamiltonian, but the adjustable function is set to $K(\eta)=1/2$. The dashed (red) curve labeled ``$C_0$'' is from Eqs.~\eqref{eq:C0hat-spin} and \eqref{eq:cGSF-C0}. The right plot shows the ISCO shift due to the spin of the orbiting test-mass. In this case the ``EOB'' curve [Eq.~\eqref{eq:cCOspinEOB}] exactly reproduces the ISCO shift computed from the Papapetrou equations (see, e.g., Appendix \ref{app:papa}; the Hamiltonian in \cite{barausse-buonanno-spinEOB} was constructed with this property). The ``$C_0$'' curve [Eq.~\eqref{eq:cCOspinC0}] agrees precisely with the exact result in the $\chi_2=0$ case. The difference between the $c^{\rm COspin}$ curves for nonzero $\chi_2$ arises from our limited knowledge of higher-order PN spin corrections.}
\end{figure*}
\subsection{\label{sec:cGSF}EOB and PN predictions for the conservative self-force ISCO shift in Kerr}
The conservative self-force ISCO shift parameter denoted $c^{\rm GSF}$ in Eq.~\eqref{eq:ISCO-expand-schematic3} is an especially interesting quantity because it is a gauge-invariant that can be calculated from self-force calculations. Barack and Sago \cite{barack-sago_isco,barack-sago-eccentricselfforce} have computed this quantity in the case of Schwarzschild, and in Ref.~\cite{favata-iscostudy} this result was compared with multiple PN-based computations of the ISCO shift.\footnote{For other comparisons of PN and GSF results, see \cite{detweiler-circselfforcePRD2008,blanchet-etal-selfforceI,blanchet-etal-selfforceII,damour-GSF,barack-damour-sago_periastron}.} Gravitational self-force results are not yet available for the Kerr spacetime, but here we explore the predictions for the conservative GSF ISCO shift in Kerr given by two PN-based calculations: the spinning-EOB approach \cite{barausse-buonanno-spinEOB} and the ISCO computed via the gauge-invariant PN ISCO condition $\hat{C}_0$ [Eq.~\eqref{eq:C0hat-spin}]. Based on the comparison study in \cite{favata-iscostudy}, these two methods are the most viable approaches for computing the ISCO in the small-mass-ratio limit.

Using the EOB ISCO frequency calculated from \cite{barausse-buonanno-spinEOB} as described above, the corresponding conservative GSF ISCO shift parameter can be computed via
\be
\label{eq:cGSF-EOB}
c^{\rm GSF}_{\rm EOB}(\chi_2) = \lim_{\eta \rightarrow 0} \frac{1}{\eta} \left[\frac{\tilde{\Omega}_{\rm EOB}(\eta, 0,\chi_2)}{\hat{\Omega}^{\rm K}(\chi_2)} -1  \right].
\ee
In the PN case a function $\tilde{\Omega}_{C_0}(\eta,\chi_1,\chi_2)$ is computed by solving for the root of Eq.~\eqref{eq:C0hat-spin} numerically. The resulting conservative GSF ISCO shift parameter is defined by
\be
\label{eq:cGSF-C0}
c^{\rm GSF}_{C_0}(\chi_2) = \lim_{\eta \rightarrow 0} \frac{1}{\eta} \left[\frac{\tilde{\Omega}_{C_0}(\eta, 0,\chi_2)}{\hat{\Omega}_{C_0}(0,0,\chi_2)} -1  \right].
\ee
Note that in this equation the denominator contains the function $\hat{\Omega}_{C_0}(0,0,\chi_2)$ rather than $\hat{\Omega}^{\rm K}(\chi_2)$. This is because the gauge-invariant PN ISCO $\tilde{\Omega}_{C_0}$ does not reduce precisely to the Kerr ISCO (although it is very close for small to moderate values of $\chi_2$; see Fig.~\ref{fig:isco-compare} and Sec.~\ref{sec:kerriscocompare}).

The resulting values for $c^{\rm GSF}_{\rm EOB}(\chi_2)$ and $c^{\rm GSF}_{C_0}(\chi_2)$ are listed in Table \ref{tab:isco-quantities} and plotted in the left-half of Fig.~\ref{fig:cEOB-PN}. Note that while the EOB curve is calibrated to the exact result in the nonspinning case, there is no expectation that it will also predict the correct ISCO shift in the spinning case. The function $K(\eta)$ will presumably need to be recalibrated when GSF results for the Kerr ISCO shift are available. To further explore the behavior of $c^{\rm GSF}_{\rm EOB}(\chi_2)$, I have varied the value of $K$ from $0$ to $4$. Figure \ref{fig:cEOB-PN} shows one of these ``uncalibrated'' choices [$K(\eta)=1/2$]. Varying $K$ over this range changes the location of the ``peak'' of $c^{\rm GSF}_{\rm EOB}(\chi_2)$. While the Barack-Sago result is no longer reproduced for other choices of $K$ (the difference with the Barack-Sago value at $\chi_2=0$ gets especially large for $K>2$), it is interesting to note that both the calibrated and uncalibrated curves approach similar values when $\chi_2 \rightarrow \pm 1$.

It will be very interesting to compare future GSF calculations of the ISCO shift in Kerr with the results shown here. Strictly speaking, the values for $c^{\rm GSF}_{C_0}(\chi_2)$ cannot be precisely compared with the ``exact'' $\chi_2\neq 0$ GSF results because the ISCO frequency in this case does not reduce precisely to the Kerr value. Still, for a large range of $\chi_2$ (as quantified in Fig.~\ref{fig:isco-compare}), an accurate comparison with future exact GSF results should still be possible. Note, in particular, that all three curves in the left-half of Fig.~\ref{fig:cEOB-PN} roughly agree for $\chi_2 \lesssim 0.2$. This is perhaps indicative that the exact GSF results will lie near those values. These predictions are likely to be most accurate for $\chi_2 \approx -1$; varying $K$ from $0$ to $4$ near this value indicates $c^{\rm GSF}_{\rm EOB}(-1) \approx 0.8\mbox{--}1.1$.
\subsection{\label{sec:cCOspin}Conservative ISCO shift due to the test-particle's spin}
It is also interesting to examine the ISCO shift parameter $c^{\rm COspin}$ [Eq.~\eqref{eq:ISCO-expand-schematic3}] originating from the spin of the point-particle. Using the EOB ISCO frequency, this quantity is calculated via
\be
\label{eq:cCOspinEOB}
c^{\rm COspin}_{\rm EOB}(\chi_2) =
\lim_{\eta \rightarrow 0} \left[ \frac{\tilde{\Omega}_{\rm EOB}(\eta,\chi_1,\chi_2)-\tilde{\Omega}_{\rm EOB}(\eta,0,\chi_2)}{\eta \chi_1 \hat{\Omega}^{\rm K}(\chi_2)} \right].
\ee
Although the quantity $c^{\rm GSF}_{\rm EOB}$ above is not exact (except for $\chi_1=0$), in this case the EOB Hamiltonian is constructed such that $c^{\rm COspin}_{\rm EOB}(\chi_2)$ is in fact the ``true'' value that would result from a calculation based on the Papapetrou-Mathisson-Dixon \cite{papapetrou-spinningI-1951,papapetrou-spinningII-1951,mathisson-1931-original,*mathisson-1931-reprint,mathisson-1937-original,*mathisson-1937-reprint,dixon-extendedbodiesI-PRSocA1970,dixon-extendedbodiesII-PRSocA1970,dixon-extendedbodiesIII-PRSocA1974,barausse-racine-buonanno-PRD2009} equations of motion.\footnote{Note that $c_{\rm EOB}^{\rm COspin}$ does not depend on $c_{\rm EOB}^{\rm GSF}$ or the choice of the adjustable function $K(\eta)$.} This was verified by an explicit calculation directly based on the Papapetrou equations (Appendix \ref{app:papa}); the two methods give identical results for $c^{\rm COspin}(\chi_2)$.
In the case of the $\hat{C}_0$ ISCO condition, we define the compact-object spin ISCO shift via
\be
\label{eq:cCOspinC0}
c^{\rm COspin}_{C_0}(\chi_2) =
\lim_{\eta \rightarrow 0} \left[ \frac{\tilde{\Omega}_{C_0}(\eta,\chi_1,\chi_2)-\tilde{\Omega}_{C_0}(\eta,0,\chi_2)}{\eta \chi_1 \hat{\Omega}_{C_0}(0,0,\chi_2)} \right],
\ee
where again the expression differs from Eq.~\eqref{eq:cCOspinEOB} because $\hat{\Omega}_{C_0}(0,0,\chi_2\neq 0)$ does not reduce to the exact Kerr ISCO.

The resulting values for $c^{\rm COspin}_{\rm EOB}(\chi_2)$ and $c^{\rm COspin}_{C_0}(\chi_2)$ are listed in Table \ref{tab:isco-quantities} and plotted in the right-half of Fig.~\ref{fig:cEOB-PN}. Note, in particular, that in the Schwarzschild case the values for $c^{\rm COspin}_{\rm EOB}$ and $c^{\rm COspin}_{C_0}$ agree precisely with each other and with the analytic calculation in Appendix \ref{subsec:Papaschw},
\be
c^{\rm COspin}_{\rm EOB}(0) = c^{\rm COspin}_{C_0}(0) = \frac{\sqrt{6}}{8} = 0.306\,186\, \ldots.
\ee
This is a remarkable result. It indicates that the gauge-invariant ISCO condition $\hat{C}_0$ not only predicts (i) the exact test-particle ISCO in the Schwarzschild case \cite{blanchetiyer3PN}, and (ii) the spin-expansion of the exact Kerr ISCO (Sec.~\ref{sec:kerriscocompare}), but it also predicts the exact shift in the Schwarzschild ISCO caused by the test-particle's spin. This shift is embodied in the (fully relativistic) Papapetrou-Mathisson-Dixon equations of motion, and it is rather unexpected that this shift could be predicted from an analysis based on the standard (nonresummed) PN equations of motion. Along with the other qualities mentioned above [and the closeness of $c_{C_0}^{\rm GSF}(0)$ to the exact Barack-Sago result], this further indicates that there is a special quality to the gauge-invariant ISCO condition in Eq.~\eqref{eq:C0hat-spin}.

In the spinning case, we see from Fig.~\ref{fig:cEOB-PN} that $c^{\rm COspin}_{C_0}(0)$ starts to deviate from the exact result as $|\chi_2|$ increases. This is due to the fact that the gauge-invariant ISCO condition $\hat{C}_0$ is limited by the number of known spin corrections in the PN equations of motion. Once higher-order spin effects have been calculated and incorporated into these calculations, it is expected that the curves labeled ``$C_0$'' in Fig.~\ref{fig:isco-compare} and the right-half of Fig.~\ref{fig:cEOB-PN} will even more closely approximate the exact results.
\section{\label{sec:conclusion}Conclusions}
The primary objective of this study was the extension of the Blanchet-Iyer \cite{blanchetiyer3PN} ISCO condition to the case of spinning, nonprecessing binaries [Eq.~\eqref{eq:C0hat-spin}]. When the test-mass limit of this condition is compared with the exact Kerr ISCO, they are found to agree up to the order to which the PN spin terms are explicitly known [cf.~Eqs.~\eqref{eq:C0kerr-smallspin} and \eqref{eq:C0spin1}, and see Fig.~\ref{fig:isco-compare}]. In addition, the conservative ISCO shifts were also computed using this ISCO condition and the spinning-EOB Hamiltonian of \cite{barausse-buonanno-spinEOB} [see Table \ref{tab:isco-quantities} and Fig.~\ref{fig:cEOB-PN}].

The ISCO shift due to the conservative gravitational self-force should eventually be compared with the exact results from self-force calculations. This will allow an extension of the study in \cite{favata-iscostudy} to the Kerr case, and will provide insight into the relative accuracies of the EOB formalism and the standard PN equations of motion. For example, in \cite{favata-iscostudy} it was found that the Blanchet-Iyer ISCO condition more accurately reproduces the Barack-Sago ISCO shift than uncalibrated EOB methods. This excellent agreement in the Schwarzschild case could be coincidental, but it would be hard to dismiss if it were also true in the Kerr case. Comparison with exact self-force results in Kerr would clarify if the standard PN equations of motion or the (uncalibrated) EOB approach can more accurately predict strong-field, finite-$\eta$ effects.

One of the most significant results of this study is that the PN ISCO condition in Eq.~\eqref{eq:C0hat-spin}---in addition to reproducing the Kerr ISCO for small spin and the Schwarzschild conservative GSF ISCO shift with good accuracy---also \emph{exactly} reproduces the ISCO shift due to the spin of the test-mass. (This agreement is truly exact only in Schwarzschild since the spin corrections in $\hat{C}_0$ are only known to quadratic order.) This provides further evidence that the ability of the $\hat{C}_0$ ISCO condition to predict strong-field results is not coincidental. However, it is somewhat mysterious as to \emph{why} this ISCO condition is able to accurately predict these strong-field effects.

In addition to explaining this agreement, future work could involve extending this study to more general orbits (such as precessing or eccentric binaries). The resulting conditions for the last stable orbit could then be compared with exact results from the Kerr spacetime.
\begin{acknowledgments}
I gratefully acknowledge Luc Blanchet and Alessandra Buonanno for their helpful comments on this manuscript. For several useful discussions I also thank the members of the relativity groups at JPL and Caltech, as well as the participants of the ``Theory meets data analysis at comparable and extreme mass ratios'' conference (Perimeter Institute, June 2010). This research was supported through an appointment to the NASA Postdoctoral Program at the Jet Propulsion Laboratory, administered by Oak Ridge Associated Universities through a contract with NASA.
\end{acknowledgments}
\appendix
\section{\label{app:spincheck}COMPARING THE PN AND KERR EXPRESSIONS FOR THE ENERGY, ANGULAR MOMENTUM, AND KEPLER RELATION}
In this appendix we examine the test-mass limit of various PN expressions, and compare them with the equivalent expressions derived from the Kerr metric. In particular, we wish to check if PN expressions using two different choices for the spin variables reduce to the same Kerr result in the test-mass limit.

The energy for circular, nonprecessing orbits is
\begin{multline}
\label{eq:E3PN}
\!\!\! \frac{E^{\rm PN}(\Omega)}{\eta M} = \! -\frac{x}{2} \! \left\{ \! 1 - x \! \left( \! \frac{3}{4} + \frac{\eta}{12}\right)
\! - x^2 \! \left( \! \frac{27}{8} -\frac{19\eta}{8} + \frac{\eta^2}{24}\right)
\right.
\\
+ x^3 \! \left[ -\frac{675}{64} + \left( \frac{34\,445}{576} -\frac{205}{96}\pi^2 \right) \eta - \frac{155}{96}\eta^2 - \frac{35}{5184}\eta^3\right]
\\
+ \frac{x^{3/2}}{M^2} \left( \frac{14}{3} S_{\ell}^{\rm c} + 2 \frac{\delta m}{M} \Sigma_{\ell}^{\rm c} \right) - \frac{x^2}{M^4} (S_{0,\ell}^{\rm c})^2
\\ \left.
+ \frac{x^{5/2}}{M^2} \left[ \left(11-\frac{61}{9}\eta \right) S_{\ell}^{\rm c} + \left(3-\frac{10}{3} \eta\right) \frac{\delta m}{M} \Sigma_{\ell}^{\rm c} \right] \right\},
\end{multline}
where the first two lines contain the nonspin terms \cite{blanchet-faye-3pn-PRD2001,blanchet35PNphase}, the third line contains the 1.5PN spin-orbit term \cite{kidder-spineffects,faye-buonanno-luc-higherorderspinI,faye-buonanno-luc-higherorderspinII} and the combined spin-spin + quadrupole monopole term (for BHs only) \cite{racine-buonanno-kidder-spinningrecoil,kidder-spineffects,poisson-quadrupolemonopoleterm-PRD1998}, and the fourth line contains the 2.5PN spin-orbit term \cite{faye-buonanno-luc-higherorderspinI,faye-buonanno-luc-higherorderspinII}.
In terms of the nonconstant-magnitude spin variables, the 2.5PN spin-orbit term can be written as \cite{faye-buonanno-luc-higherorderspinII}
\be
+ \frac{x^{5/2}}{M^2} \left[ \left(13-\frac{49}{9}\eta \right) S_{\ell}^{\rm nc} + \left(5-\frac{8}{3} \eta\right) \frac{\delta m}{M} \Sigma_{\ell}^{\rm nc} \right],
\ee
while the 1.5PN and 2PN spin terms keep the same form but with ``c'' replaced by ``nc''.
In the test-mass limit $E^{\rm PN}$ reduces to
\begin{multline}
\label{eq:Epntestmass}
\frac{E^{\rm PN}(\Omega)}{m_1} = -\frac{1}{2} x \left[ 1 -\frac{3}{4} x +\frac{8}{3} \chi_2 x^{3/2}
\right. \\ \left.
- x^2 \left( \frac{27}{8} + \chi_2^2 \right) + 8 \chi_2 x^{5/2} -\frac{675}{64}x^3 \right],
\end{multline}
where $\chi_2$ can be either $\chi_2^{\rm nc}$ or $\chi_2^{\rm c}$.

The total energy of a point-mass in the Kerr spacetime is given in terms of $w_{\rm BL}$ in Eq.~\eqref{eq:Ekerr}. [Recall that $\tilde{E}$ includes the particle's rest mass, so the orbital energy is $\tilde{E}-1$.] Substituting Eq.~\eqref{eq:wBL} and expanding in $X$ yields
\begin{multline}
\tilde{E}-1 = -\frac{X}{2} \! \left\{ 1 -\frac{3}{4} X + \frac{8}{3} \chi_2^{\rm K} X^{3/2} - X^2 \! \left[ \frac{27}{8} + (\chi_2^{\rm K})^2 \! \right]
\right. \\
+ 8 \chi_2^{\rm K} X^{5/2} + X^3 \left[ -\frac{675}{64} - \frac{65}{18} (\chi_2^{\rm K})^2 \right]
\\ \left.
+ 27 \chi_2^{\rm K} X^{7/2} + O(X^4)  \right\},
\end{multline}
which agrees with Eq.~\eqref{eq:Epntestmass} to the expected order.

The orbital angular momentum (specialized to equatorial orbits) is given by Eqs.~(6.10) and (7.10) of \cite{faye-buonanno-luc-higherorderspinII},
\begin{multline}
\label{eq:Ldotl}
{\mathbf L} \cdot {\bm \ell} = \frac{\eta M^2}{x^{1/2}} \left\{ 1 + x\left(\frac{3}{2} + \frac{\eta}{6}\right) + x^2 \left( \frac{27}{8} -\frac{19\eta}{8} + \frac{\eta^2}{24} \right) \right. \\
 + \frac{x^{3/2}}{M^2} \left( -\frac{35}{6} S_{\ell}^{\rm c} - \frac{5}{2}\frac{\delta m}{M} \Sigma_{\ell}^{\rm c} \right)
\\ \left.
+ \frac{x^{5/2}}{M^2} \left[ \left( \! -\frac{77}{8} + \frac{427}{72}\eta\right) S_{\ell}^{\rm c} + \! \left( \! -\frac{21}{8} + \frac{35}{12}\eta \right) \! \frac{\delta m}{M} \Sigma_{\ell}^{\rm c} \right] \! \right\} \!,
\end{multline}
with the last two lines replaced by the following expression in terms of the ``nc'' spin variables:
\begin{multline}
+ \frac{x^{3/2}}{M^2} \left( -\frac{23}{6} S_{\ell}^{\rm nc} - \frac{3}{2}\frac{\delta m}{M} \Sigma_{\ell}^{\rm nc} \right)
\\ \left.
+ \frac{x^{5/2}}{M^2} \left[ \left( \! -\frac{77}{8} + \frac{259}{72}\eta\right) S_{\ell}^{\rm nc} + \! \left( \! -\frac{33}{8} + \! \frac{7}{4}\eta \right) \! \frac{\delta m}{M} \Sigma_{\ell}^{\rm nc} \right] \! \right\} \! .
\end{multline}
Note that the 2PN spin(1)-spin(2) term is zero \cite{kidder-spineffects}, but the 2PN quadrupole-monopole contribution has not been computed. The 3PN nonspin terms are given in general form in \cite{deAndrade-blanchet-faye-CQG2001}, but have not been specified to circular orbits. Also note that the spin-orbit terms in ${\mathbf L}$ differ even at 1.5PN order when one switches spin variable.\footnote{The total angular momentum ${\mathbf J} = {\mathbf L} + {\mathbf S}_1/c + {\mathbf S}_2/c$ is a constant vector (up to 2PN order) that does not depend on the choice of spin variable. Since the individual spins contribute a 0.5PN correction to the total angular momentum, the 1PN corrections in the relations between spin variables [Eqs.~\eqref{eq:relatespins}] shift some terms into (or out of) the 1.5PN piece of ${\mathbf L}$.}
In the test-mass limit, these expressions reduce to
\begin{multline}
\label{eq:Lpntestmass2}
\frac{{\mathbf L} \cdot {\bm \ell}}{m_1} = \frac{m_2}{x^{1/2}} \left\{ 1 + \frac{3}{2} x - \frac{10}{3} \chi_2^{\rm c} x^{3/2}
\right. \\ \left.
+ x^2 \left[ \frac{27}{8} + C_{\rm QM}^{\rm c} (\chi_2^{\rm c})^2 \right] -7\chi_2^{\rm c} x^{5/2} + O(x^3) \right\},
\end{multline}
\begin{multline}
\label{eq:Lpntestmass1}
\frac{{\mathbf L} \cdot {\bm \ell}}{m_1} = \frac{m_2}{x^{1/2}} \left\{ 1 + \frac{3}{2} x - \frac{7}{3} \chi_2^{\rm nc} x^{3/2}
\right. \\ \left.
+ x^2 \left[ \frac{27}{8} + C_{\rm QM}^{\rm nc} (\chi_2^{\rm nc})^2 \right] -\frac{11}{2} \chi_2^{\rm nc} x^{5/2} + O(x^3) \right\},
\end{multline}
where the constants $C_{\rm QM}^{\rm c}$ and $C_{\rm QM}^{\rm nc}$ have not been explicitly computed.

The orbital angular momentum of a test-particle in Kerr is \cite{bpt}
\be
\label{eq:Lkerr}
\tilde{L} \equiv \frac{L}{m_1} = \frac{\sign(\chi_2^{\rm K}) m_2}{\sqrt{w_{\rm BL}}} \frac{[1-2\chi_2^{\rm K} w_{\rm BL}^{3/2} + (\chi_2^{\rm K})^2 w_{\rm BL}^2]}{\sqrt{1-3w_{\rm BL}+2\chi_2^{\rm K} w_{\rm BL}^{3/2}}}.
\ee
Substituting Eq.~\eqref{eq:wBL} and expanding in $X$ yields
\begin{multline}
\label{eq:Lkerrexpand}
\tilde{L} = \frac{\sign(\chi_2^{\rm K}) m_2}{X^{1/2}} \left\{ 1 + \frac{3}{2} X - \frac{10}{3} \chi_2^{\rm K} X^{3/2}
\right. \\
+ X^2 \left[ \frac{27}{8} + (\chi_2^{\rm K})^2 \right] - 7 \chi_2^{\rm K} X^{5/2} + X^3 \left[ \frac{135}{16} + \frac{26}{9} (\chi_2^{\rm K})^2 \right]
\\ \left.
- \frac{81}{4} \chi_2^{\rm K} X^{7/2} +O(X^4) \right\}.
\end{multline}
Here we see that the Kerr angular momentum agrees with the test-mass limit of the PN expression only if we identify $\chi_2^{\rm K}$ with $\chi_2^{\rm c}$. Note also that Eq.~\eqref{eq:Lkerrexpand} provides the test-mass limit of the previously unknown 2PN and 3PN pieces of Eq.~\eqref{eq:Ldotl}.

We also check for agreement between the PN and Kerr versions of the Keplerian relationship (see also Appendix B of \cite{faye-buonanno-luc-higherorderspinI}). The PN relation is given in Eqs.~\eqref{eq:gamma-x-spin} and \eqref{eq:gamma-x-spin-nc}. In the test-mass limit it reduces to
\begin{multline}
\label{eq:gamma-testmass}
\gamma \rightarrow  \frac{m_2}{r_H} = x \left[ 1 + x + \frac{2}{3} \chi_2 x^{3/2} + x^2 \left( 1- \frac{\chi_2^2}{2} \right)
\right. \\ \left.
+ \frac{4}{3} \chi_2 x^{5/2} + x^3 + O(\chi_2^2 x^3) \right],
\end{multline}
where $\chi_2$ can be either spin variable. Note that the PN radial coordinate used in the main text refers to harmonic coordinates (here denoted $r_H$).
To derive the Kerr-analog of this expression we first need the relationship between Boyer-Lindquist and harmonic coordinates \cite{tagoshi-ohashi-owen-2.5SOterm,cook-scheel=PRD1997},
\bs
\label{eq:BL-harmonic}
\begin{align}
x_H + i y_H &= (r_{\rm BL} - m_2 + i \chi_2^{\rm K} m_2) e^{i\varphi} \sin\theta_{\rm BL}, \\
z_H &= (r_{\rm BL} - m_2) \cos\theta_{\rm BL}, \\
r_H^2 &= x_H^2+y_H^2+z_H^2 = (r_{\rm BL}-m_2)^2 \nonumber
\\  &+ (\chi_2^{\rm K} m_2)^2 \sin^2\theta_{\rm BL}.
\end{align}
\es
Specializing to the equatorial plane ($\theta_{\rm BL}=\pi/2$) and defining $w_H\equiv m_2/r_H$, we have the relationship
\be
\label{eq:wH}
w_H = \frac{w_{\rm BL}}{\sqrt{(1-w_{\rm BL})^2 + (\chi_2^{\rm K} w_{\rm BL})^2}}.
\ee
Substituting Eq.~\eqref{eq:wBL} for $w_{\rm BL}$ and series expanding in $X$ yields
\begin{multline}
\label{eq:wH-X}
w_H = X \left\{ 1 + X + \frac{2}{3} \chi_2^{\rm K} X^{3/2} + X^2 \left[ 1 - \frac{(\chi_2^{\rm K})^2}{2} \right]
\right. \\
+ \frac{4}{3} \chi_2^{\rm K} X^{5/2} + X^3 \left[ 1- \frac{17}{18} (\chi_2^{\rm K})^2 \right]
\\ \left.
+ X^{7/2} \left[ 2 \chi_2^{\rm K} - (\chi_2^{\rm K})^3 \right] + O(X^4) \right\},
\end{multline}
which agrees with Eq.~\eqref{eq:gamma-testmass} to the expected order.

\section{\label{app:papa}SPINNING TEST-PARTICLE ISCO SHIFT DERIVED FROM THE PAPAPETROU- MATHISSON-DIXON EQUATIONS}
In this appendix I discuss how to compute the ISCO for a spinning test-particle directly from the Papapetrou-Mathisson-Dixon equations (rather than from the EOB formalism of \cite{barausse-buonanno-spinEOB}). The results for the ISCO shift parameter derived below agree exactly with the results discussed in Sec.~\ref{sec:cCOspin}, Table \ref{tab:isco-quantities}, and the right plot of Fig.~\ref{fig:cEOB-PN}. This provides further confirmation of the validity of the Hamiltonian derived in \cite{barausse-racine-buonanno-PRD2009,barausse-buonanno-spinEOB}. Previous examinations of the ISCO of a spinning test-particle are given in \cite{suzuki-maeda-COspinISCO,tanaka-mino-sasaki-shibata-PRD1996-spinningparticlePN} (see also \cite{suzuki-maeda-PRD1997-chaos-schw}). The results here are more explicit, exact numerical values are given (Table \ref{tab:isco-quantities}), and a fully analytic examination in Schwarzschild is presented.

\subsection{\label{subsec:Papaeqns}Papapetrou-Mathisson-Dixon equations for equatorial, nonprecessing orbits}
Saijo et al.~\cite{saijo-maeda-shibata-mino-spinningparticleplunge-equatorial} have explicitly derived the equations of motion of a spinning particle in the equatorial plane ($\theta=\pi/2$) of a Kerr BH. For a particle with spin angular momentum ${\bm S}_1 = s m_1 \hat{\bm z}$ aligned with the BH's spin (${\bm S}_2 = a m_2 \hat{\bm z}$) and the orbital angular momentum $L_z$, the spin vectors remain constant and the equations of motion take a form similar to the Kerr geodesic equations [Eqs.~(2.19)-(2.25) of \cite{saijo-maeda-shibata-mino-spinningparticleplunge-equatorial}]:
\bs
\label{eq:papapetroueqns}
\begin{align}
\label{eq:dtdtau}
\Sigma_s \Lambda_s \frac{dt}{d\tau} &= \! a \left( 1 + \frac{3m_2 s^2}{r \Sigma_s} \right) \! \left[ \tilde{J}_z - (a + s) \tilde{E} \right] \! + \! \frac{r^2+a^2}{\Delta}P_s, \\
\label{eq:dphidtau}
\Sigma_s \Lambda_s \frac{d\varphi}{d\tau} &= \left( 1 + \frac{3m_2 s^2}{r \Sigma_s} \right) \left[ \tilde{J}_z - (a + s) \tilde{E} \right] + \frac{a}{\Delta}P_s, \\
\label{eq:drdtau}
\Sigma_s \Lambda_s \frac{dr}{d\tau} &= \pm \sqrt{R_s}, \;\;\; \text{where}
\end{align}
\begin{align}
\Sigma_s &= r^2 \left( 1 - \frac{m_2 s^2}{r^3} \right), \\
\Lambda_s &= 1- \frac{3 m_2 s^2 r [\tilde{J}_z - (a+s)\tilde{E} ]^2}{\Sigma_s^3}, \\
R_s &=P_s^2 - \Delta \left\{ \frac{\Sigma_s^2}{r^2} + [ \tilde{J}_z - (a+s)\tilde{E} ]^2 \right\}, \\
P_s &= \left[ (r^2+a^2) + a s \left( 1 + \frac{m_2}{r} \right) \right] \tilde{E} - \left( a + s \frac{m_2}{r} \right) \tilde{J}_z, \\
\Delta &=r^2-2m_2 r + a^2,
\end{align}
\es
where $(t,r,\theta,\varphi)$ are Boyer-Lindquist coordinates\footnote{Note that in the rest of this paper $r$ denotes the harmonic radial coordinate. Also, to maintain \emph{some} notational consistency, I continue to denote the central BH mass by $m_2$ and the test-particle's mass by $m_1$; in most of the literature on the Papapetrou equations these quantities are denoted $M$ and $\mu$ respectively.}, $\tau$ is the particle's proper time, and the conserved energy $\tilde{E}\equiv E/m_1$ and total angular momentum $\tilde{J}_z \equiv J_z/m_1$ are given in Eqs.~(2.10) of \cite{saijo-maeda-shibata-mino-spinningparticleplunge-equatorial}.

Note that the function $R_s$ can be rewritten in the form
\be
R_s = B(r) [\tilde{E}-\tilde{E}_1(r,\tilde{J}_z)][\tilde{E}-\tilde{E}_2(r,\tilde{J}_z)],
\ee
where the roots $\tilde{E}_{1,2}$ of $R_s=0$ are found by solving [see also Eq.~(2.26) of \cite{saijo-maeda-shibata-mino-spinningparticleplunge-equatorial}]
\be
\alpha \tilde{E}^2 - 2 \beta \tilde{E} + \gamma =0, \;\;\;\; \text{with}
\ee
\begin{align}
\alpha &= \left[ (r^2 + a^2) + a s \left( 1 + \frac{m_2}{r} \right) \right]^2 - \Delta (a + s)^2 , \\
\beta &= \! \left\{ \! \left( a + s \frac{m_2}{r} \right) \! \left[ \! (r^2 + a^2) \! + a s \left( \! 1 \! + \! \frac{m_2}{r} \right) \! \right] \! - \! \Delta (a \! + \! s) \! \right\} \! \tilde{J}_z, \\
\gamma &= \left( a + s \frac{m_2}{r} \right)^2 \tilde{J}_z^2 - \Delta \left[ r^2 \left( 1 - \frac{M s^2}{r^3} \right)^2 + \tilde{J}_z^2 \right].
\end{align}
Here $(\alpha, \beta, \gamma)$ are not to be confused with any quantities defined earlier in this paper. The solution
\be
\label{eq:E1Veff}
\tilde{E}_1 \equiv V_{\rm eff} = \frac{\beta + \sqrt{\beta^2 - \alpha \gamma}}{\alpha}
\ee
corresponds to an effective potential for the particle motion.\footnote{Eq.~(2.27) of \cite{saijo-maeda-shibata-mino-spinningparticleplunge-equatorial} has the wrong sign in front of the $\alpha \gamma$ term.} Here we have taken the positive square root to ensure that the particle energy $\tilde{E}=\tilde{E}_1 \rightarrow 1$ when $r \rightarrow \infty$ (in contrast to the negative root, for which $\tilde{E}_2 \rightarrow -1$). This allows us to rewrite the equation for the radial motion in the form
\be
\label{eq:dotrsq}
\dot{r}^2 = A(r,\tilde{E}, \tilde{J}_z) [\tilde{E}-V_{\rm eff}(r,\tilde{J}_z)],
\ee
where, for this appendix \emph{only}, an overdot means $d/d\tau$.
The explicit forms for $A$ and $B$ can be inferred from the above equations but are not needed for the remainder of the analysis.

\subsection{\label{subsec:Papaisco}General solution for the ISCO of a spinning particle}
The conditions for circular orbits (defined as orbits with constant $r$) are that both $\dot{r}$ and $\ddot{r}$ vanish. By differentiating Eq.~\eqref{eq:dotrsq} and dividing by $\dot{r}$,
\be
\label{eq:ddotr}
\ddot{r} = \frac{1}{2} \left[ (\tilde{E}-V_{\rm eff}) \frac{\partial A}{\partial r} - A \frac{\partial V_{\rm eff}}{\partial r} \right],
\ee
we see that the conditions for circular orbits are equivalent to
\be
\tilde{E}=V_{\rm eff}(r,\tilde{J}_z) \;\;\;\; \text{and} \;\;\;\; \frac{\partial V_{\rm eff}(r,\tilde{J}_z)}{\partial r} =0.
\ee
To ensure that circular orbits are stable, we require that under a small radial perturbation of a circular orbit, $r_0 \rightarrow r_0 + \delta r$, the particle is accelerated back to its initial configuration. Such a condition is equivalent to demanding that the perturbed coordinate acceleration satisfy $\ddot{\delta r} = - \tilde{\omega}_0^2 \delta r$ with $\tilde{\omega}_0^2>0$, where $\tilde{\omega}_0$ is the radial oscillation frequency about the unperturbed orbit $r_0$ [this is equivalent to the analysis in Eqs.~\eqref{eq:linearpert}\mbox{--}\eqref{eq:C0cond} above]. In this case $\tilde{\omega}_0$ is found by linearizing Eq.~\eqref{eq:ddotr} about the circular orbit $r_0$. Computing $\partial \ddot{r}/\partial r$ and evaluating along the unperturbed circular orbit yields
\be
\tilde{\omega}_0^2 = - \left. \frac{\partial \ddot{r}}{\partial r} \right|_0 = \frac{A}{2} \frac{\partial^2 V_{\rm eff}}{\partial r^2}.
\ee
The ISCO is found from the equality $\tilde{\omega}_0^2=0$ (note that $A$ is nonzero for physically relevant parameter values).

To evaluate the ISCO frequency, we first solve the algebraic system of equations
\be
\frac{\partial V_{\rm eff}(r,\tilde{J}_z)}{\partial r} =0 \;\;\;\; \text{and} \;\;\;\; \frac{\partial^2 V_{\rm eff}(r,\tilde{J}_z)}{\partial r^2} =0
\ee
for the ISCO values of $(r,\tilde{J}_z)$.
This is done numerically, specifying $a=\chi_2 m_2$, $s=\chi_1 q m_2$, $m_2=1$, and using $r=r_{\rm isco}^{\rm K}$ [Eq.~\eqref{eq:risco-kerr}] and $\tilde{J}_z = \tilde{L}(r_{\rm isco}^{\rm K})$ [Eq.~\eqref{eq:Lkerr}] as initial guesses for the solution. The resulting values $(r_0,\tilde{J}_0)$ are then used to determine the ISCO energy $\tilde{E}_0=V_{\rm eff}(r_0,\tilde{J}_0)$. The ISCO angular frequency is then found by substituting these quantities into
\be
\label{eq:Omegapapa}
\Omega \equiv \frac{d\varphi/d\tau}{dt/d\tau}
\ee
using Eqs.~\eqref{eq:dtdtau} and \eqref{eq:dphidtau}. This procedure allows the ISCO frequency to be computed as a function of $(q,\chi_1,\chi_2)$.

The ISCO shift parameter $c^{\rm COspin}(\chi_2)$ is computed as in Eq.~\eqref{eq:cCOspinEOB}. Note that in this case $c^{\rm GSF}$ evaluates to zero (as expected) and converting variables from $(q,m_2\Omega)$ to $(\eta,M\Omega)$ does not affect the value of $c^{\rm COspin}(\chi_2)$ [see Eq.~\eqref{eq:ISCO-expand-schematic3}]. The resulting values for $c^{\rm COspin}(\chi_2)$ are identical to those listed in Table \ref{tab:isco-quantities} under $c^{\rm COspin}_{\rm EOB}(\chi_2)$.

\subsection{\label{subsec:Papaschw}Analytic analysis of the ISCO in the Schwarzschild, small-spin limit}
It is instructive to reexamine the above analysis of the ISCO, specializing to Schwarzschild ($a=0$) and small spin ($s/m_2\ll 1$).\footnote{Since $s/m_2=\chi_1 q$, the small-spin limit is still quite accurate for EMRIs since $q\ll 1$ even if $\chi_1 \sim 1$.} Keeping terms linear in $s$, Eqs.~\eqref{eq:papapetroueqns} reduce to
\bs
\begin{align}
\frac{dt}{d\tau} &= \frac{\tilde{E}}{1-\frac{2 m_2}{r}} - \frac{s m_2 \tilde{J}_z}{r^3 \left( 1- \frac{2m_2}{r} \right) } + O(s^2), \\
\frac{d\varphi}{d\tau} &= \frac{\tilde{J}_z}{r^2} - \frac{s \tilde{E}}{r^2} + O(s^2), \\
\left( \frac{dr}{d\tau} \right)^2 \!\!\! &= \! \tilde{E}^2 \! - \! [V_{\rm eff}^{\rm schw}(r,\tilde{J}_z)]^2 \! + 2s \frac{\tilde{E} \tilde{J}_z}{r^2} \! \left( \! 1 \! - \! \frac{3 m_2}{r} \! \right) \! + \! O(s^2),
\end{align}
\es
\be
\text{where} \;\;\;\; (V_{\rm eff}^{\rm schw})^2 \equiv \left( 1-\frac{2 m_2}{r} \right) \left( 1 + \frac{\tilde{J}_z^2}{r^2} \right)
\ee
is the effective potential for Schwarzschild. Setting $\dot{r}^2 =0$ yields a quadratic equation for $\tilde{E}$ which, when solved and expanded in $s$, yields
\be
\label{eq:Veffspin}
\tilde{E} \equiv V_{\rm eff}^{\rm schw,\,spin} = V_{\rm eff}^{\rm schw} - \frac{s \tilde{J}_z}{r^2} \left( 1- \frac{3m_2}{r} \right) + O(s^2).
\ee
Note that this is equivalent to the $O(s)$ expansion of Eq.~\eqref{eq:E1Veff} with $a=0$.

For circular orbits we solve the condition $\partial V_{\rm eff}^{\rm schw,\,spin}/\partial r=0$ for $\tilde{J}_z = \tilde{J}_z^{\rm schw} + s \delta \hat{J}_z + O(s^2)$, yielding the angular momentum for circular orbits,
\be
\label{eq:Jzcirctot}
\tilde{J}_z^{\rm circ} = \frac{r\sqrt{m_2}}{\sqrt{r - 3 m_2}} + \frac{s}{2} \frac{(r-2m_2)(2r-9m_2)}{\sqrt{r} (r-3m_2)^{3/2}} + O(s^2).
\ee
Substituting into Eq.~\eqref{eq:Veffspin} and expanding in $s$ yields the energy along circular orbits,
\be
\label{eq:Ecirctot}
\tilde{E}^{\rm circ} = \frac{r-2m_2}{\sqrt{r(r-3m_2)}} - \frac{s}{2r} \left( \frac{m_2}{r-2m_2} \right)^{3/2} + O(s^2).
\ee

To determine the ISCO we compute $\partial^2 V_{\rm eff}^{\rm schw,\,spin}/\partial r^2=0$, substitute Eq.~\eqref{eq:Jzcirctot} for $\tilde{J}_z$, expand to $O(s)$, and solve for $r= 6 m_2 + s \delta \hat{r} + O(s^2)$. The resulting ISCO radius is
\be
\label{eq:risco-papa}
r_{\rm isco} = 6 m_2 - 2 s \sqrt{\frac{2}{3}} + O(s^2).
\ee
Substituting this result into Eqs.~\eqref{eq:Ecirctot} and \eqref{eq:Jzcirctot} gives the energy and angular momentum at the ISCO,
\bs
\label{eq:EJcircpapa}
\begin{align}
\tilde{E}^{\rm isco} &= \frac{2\sqrt{2}}{3} -\frac{\sqrt{3}}{108} \frac{s}{m_2} + O(s^2), \\
\tilde{J}_z^{\rm isco} &= 2\sqrt{3} m_2 + \frac{\sqrt{2}}{3} s + O(s^2).
\end{align}
\es

To compute the ISCO frequency we expand Eq.~\eqref{eq:Omegapapa},
\be
\Omega = \frac{\tilde{J}_z}{r^2 \tilde{E}} \! \left( 1-\frac{2m_2}{r} \right) - \frac{s (r-2m_2)}{r^3} \! \left( 1-\frac{m_2 \tilde{J}_z^2}{r^3 \tilde{E}^2} \right) + O(s^2),
\ee
substitute Eqs.~\eqref{eq:risco-papa} and \eqref{eq:EJcircpapa}, and expand to $O(s)$,
\be
m_2 \Omega_{\rm isco} = 6^{-3/2} + \frac{1}{48}\frac{s}{m_2} + O(s^2).
\ee
Multiplying by $M/m_2$ and using $s=\chi_1 q m_2$ and $q=\eta + O(\eta^2)$, we can write the shift in the ISCO as
\be
M\Omega_{\rm isco} = 6^{-3/2} [ 1 + \eta + c^{\rm COspin}_{\rm schw} \chi_1 \eta + O(\eta^2) ],
\ee
where the $O(\eta)$ term would combine with the conservative GSF shift (not computed here), and
\be
\label{eq:cCOspin-papaschw}
c^{\rm COspin}_{\rm schw} = \frac{\sqrt{6}}{8} = 0.306\,186\,217\,847\ldots .
\ee
This agrees with the Schwarzschild value found from three separate calculations via the EOB Hamiltonian [Eq.~\eqref{eq:cCOspinEOB}], the gauge-invariant ISCO condition [Eq.~\eqref{eq:cCOspinC0}], and the numerical evaluation of the Papapetrou equations (Appendix \ref{subsec:Papaisco} above).
\bibliography{text_favata_kerrISCOshift_v2b}
\end{document}